\documentclass[a4paper, twoside, 12pt, fleqn] {article} %{report}

\usepackage{latexsym}           % permet l'utilisation de la police des
\usepackage{amsmath}            %
\usepackage{exscale}            % fournit des versions de taille
\usepackage{amsfonts}
\usepackage{amsthm}
\usepackage{amssymb}
\usepackage{ifthen}

\newlength{\intlen}
\newcommand{\integral}[2][]{%
    \settowidth{\intlen}{$\int_{#1}$}
    \int_{#1}\hspace{-1\intlen}
    \settowidth{\intlen}{$\int$}
    \hspace{0.8\intlen} {#2}\,}
\newcommand{\integralbis}[3]{%
    \settowidth{\intlen}{$\int_{#1}^{#2}$}
    \int_{#1}^{#2}\hspace{-1\intlen}
    \settowidth{\intlen}{$\int$}
    \hspace{0.8\intlen} {#3}\,}

\newlength{\Inta}
\newlength{\Intb}
\newcommand{\Int}[2][]{\integral[#1]{#2}
    \settowidth{\Inta}{$\int_{#1}$}
    \settowidth{\Intb}{$\integral[#1]{#2}$}
    \ifthenelse{\lengthtest{\Inta >
\Intb}}{\addtolength{\Inta}{-1\Intb}\hspace{1\Inta}\,}{}}
\newcommand{\Integral}[3]{\integralbis{#1}{#2}{#3}%
    \settowidth{\Inta}{$\int_{#1}^{#2}$}
    \settowidth{\Intb}{$\integralbis{#1}{#2}{#3}$}
    \ifthenelse{\lengthtest{\Inta >
\Intb}}{\addtolength{\Inta}{-1\Intb}\hspace{1\Inta}\,}{}}

\newcommand{\B}[1]{\mathbf{#1}}
\newcommand{\avg}[1]{\left\langle #1 \right\rangle} %{\langle #1 \rangle}
\renewcommand{\d}{\mathrm{d}}
\newcommand{\p}{\partial}
\newcommand{\e}{\mathrm{e}} % exponentielle
\newcommand{\0}{\mathbf{0}}

\renewcommand{\r}{\mathbf{r}}
\newcommand{\y}{\mathbf{y}}
\renewcommand{\k}{\mathbf{k}}
\newcommand{\q}{\mathbf{q}}
\newcommand{\KA}{{\kappa_A}}	% enclosed in curly brackets to avoid pb of
\newcommand{\KB}{{\kappa_B}}
\newcommand{\vSR}{{v_{\text{SR}}}}
\newcommand{\LA}{{\Lambda_A}}
\newcommand{\LB}{{\Lambda_B}}

\newcommand{\Fr}{{F^{\text{R}}}}

\begin{document}
	
\title{\bf Microscopic origin of universality in Casimir forces}
\author{Pascal R. Buenzli {\normalsize and} Philippe A. Martin\\ Institute
    of Theoretical Physics\\ Swiss Federal Institute of Technology
    Lausanne\\ CH-1015, Lausanne EPFL, Switzerland}

\date{\today}

\maketitle

\begin{abstract}
    The microscopic mechanisms for universality of Casimir forces between
    macroscopic conductors are displayed in a model of classical charged
    fluids. The model consists of two slabs in empty space at distance $d$
    containing classical charged particles in thermal equilibrium (plasma,
    electrolyte).  A direct computation of the average force per unit
    surface yields, at large distance, the usual form of the Casimir force
    in the classical limit (up to a factor 2 due to the fact that the model
    does not incorporate the magnetic part of the force). Universality
    originates from perfect screening sum rules obeyed by the microscopic
    charge correlations in conductors.  If one of the slabs is replaced by
    a macroscopic dielectric medium, the result of Lifshitz theory for the
    force is retrieved. The techniques used are Mayer expansions and
    integral equations for charged fluids.
\end{abstract}

\vskip 0.5cm
\noindent {\bf KEYWORDS~:} Casimir forces, classical charged fluid,
universality, Mayer expansion
\vskip 0.5cm
\noindent {\bf PACS numbers~:} {05.20.-y, 61.20.-p}
\vskip 0.5cm
\noindent {\it Corresponding author~:} Pascal Buenzli,
e-mail~: Pascal.Buenzli@epfl.ch

%\tableofcontents
\pagebreak

\section{Introduction}
It is well known that the fluctuation-induced forces between macroscopic
conductors have an universal character~: they only depend on the shapes of
the bodies, but not on their material constitution. This observation
originates from the celebrated paper of H.  B. G. Casimir calculating the
force between two parallel metallic plates due to the fluctuations of the
quantum electromagnetic field in vacuum at zero temperature. The literature
produced since then is so vast that we will only quote in the sequel a
number of papers relevant to our purpose. Many references can be found for
instances in the books and reviews \cite{milonni} \cite{mostepanenko}
\cite{plunien} \cite{poincare}.

Concerning the extension of Casimir's result to non zero temperature $T$,
Balian and Duplantier provide the general form of the free energy in
presence of ideal conductors of arbitrary shapes
\cite{balian-duplantier}. The theory of Lifshitz and coworkers generalizes
the calculations to dielectric bodies caracterized by their dielectric
fonctions \cite{lifshitz1}, \cite{lifshitz2} (see also \cite{schwinger2}).
The ideal conductor
situation can be recovered from the latter theory by letting the dielectric
constants tend to infinity.  From these studies one can obtain the
asymptotic behaviour of the attractive force between two planar conductors
at distance $d$ at high temperature (or equivalently at large separation
$d$)
\footnote{The relevant dimensionless parameter is $d \;k_{B}T/\hbar c$, $c$
    is the speed of light, $\hbar$ the Planck constant, $k_{B}$ the
    Boltzmann constant.}
\begin{equation}
    f\sim \frac{k_{B}T\zeta(3)}{4\pi d^{3}}, \quad d\to\infty
    \label{0.1}
\end{equation}
where $\zeta$ is the Riemann $\zeta$-function. In this regime the force is
exclusively due to thermal fluctuations and the result may be called
classical since it does not depend on Planck's constant. In the above
mentioned theories the conductors are treated at the level of macroscopic
physics. In fact they are represented by surfaces, called ideal conductors,
on which the electromagnetic field has to satisfy the metallic boundary
conditions.  The purpose of this work is to gain an understanding of the
microscopic mechanisms in the conductor that lead to the universality of
the force (\ref{0.1}).

To this end we analyse a simple model where the conductors are described in
fully microscopic terms. The conductors consist of two slabs at distance
$d$ containing fluids of classical charges (e.g. classical electrolytes or
plasmas).  The slabs are globally neutral but their material composition
(charges and masses of the particles) can be different. The space external
to the slabs is empty. The system of the two slabs, considered as a whole,
is at thermal equilibrium with a Gibbs weight that includes pairwise
interactions between all the particles, consisting of Coulomb potentials
plus short-range repulsions. In this setting we present an exact
computation of the asymptotic behaviour of the average force per unit
surface between the two (infinitely-thick) slabs giving
\begin{equation}
    \langle f\rangle\sim \frac{k_{B}T\zeta(3)}{8\pi d^{3}}, \quad
    d\to\infty.
\label{0.1a}
\end{equation}
It is also checked that (\ref{0.1a}) still holds for slabs of finite
thickness (appendix A).

One may notice that the usual approaches consider the fluctuating
electromagnetic fields as primary objects. The thermal fluctuations of
these fields originate from the fact that they are in equilibrium with the
matter constituting the conductors, but as a consequence of universality,
the microscopic degrees of freedom of the charges in the conductors do not
need to be explicitly incorporated in the description.  Here we adopt
another point of view~: we start from the thermal configurational
fluctuations of the charges to provide a direct calculation of the average
force without recourse to the field statistics.  Then the origin of
universality can be traced back to the specific sum rules obeyed by the
correlations of Coulombic matter \cite{martin}.

Universal properties of a variety of classical models of conductor have
been studied in \cite{forrester-janco-tellez} \cite{janco-tellez}. In
\cite{forrester-janco-tellez} the authors consider a statistical mechanical
system of charges confined to a plane at distance $d$ of another ideal
planar conductor and establish the result (\ref{0.1a}). In
\cite{janco-tellez}, they show that replacing the above ideal conductor by
fluctuating charges does not alter (\ref{0.1a}).

The value (\ref{0.1}) arising from the electromagnetic field fluctuations
calculations is twice larger than that obtained in the purely electrostatic
models considered here, as well as in \cite{forrester-janco-tellez}
\cite{janco-tellez}. This point has been the subject of several discussions
in the literature, in particular in \cite{schwinger1} \cite{schwinger2}. In
\cite{schwinger1} Schwinger performs a calculation of the Casimir force
mediated by scalar photons (corresponding to the sole electric degree of
freedom of an electromagnetic wave) leading to the result (\ref{0.1a}).  In
\cite{schwinger2} the authors show that taking the magnetic degree of
freedom of the field property into account multiplies the expression
(\ref{0.1a}) by $2$. In its very formulation our model does not include the
magnetic part of the Lorentz force induced by fluctuating currents in the
conductors, whose effect has the same magnitude as that of the Coulomb
force. Although such purely electrostatic models of conductors do not
account for the physically correct value of the force at large distance,
they already nicely reveal the microscopic mechanisms occurring in
conductors that guarantee its universality.

The calculation of the force requires the knowledge of the charge
correlation function across the two slabs separation, which is the main
object of our study. At large separation it remarkably factorizes into
three parts. There is a first factor independent of the slabs' material
constitution and two other factors, each solely associated to one of the
conductors. More precisely, the latter factors involve the charge density
of the screening cloud induced by a charge located at the boundary of a
single conductor in empty space. Then the universality of the force results
from the perfect screening sum rule that holds in any conducting phase.

In this work we use the technique of Mayer expansion and integral equations
for charged fluids.  In section 2, we specify the system under study and
express the force by unit surface between two infinitely extended slabs in
terms of the microscopic charge correlation between them, taking the
existence of the thermodynamic limit for granted.

The general formalism used is recalled in section \ref{sec:mayer}~: the
charge correlation function is written in terms of the Ursell function
subjected to a Mayer expansion. The prototype graphs entering in this
expansion involve screened Coulomb bonds resulting from chain summations
(Debye-H\"uckel mean-field potential) and density weights at vertices
\cite{meeron} \cite{aqua-cornu-mayer}. The weights are the exact
inhomogeneous densities that have to be self-consistently determined from
the first BGY equation. We do not treat here the full self-consistent
problem because it turns out that the detailed structure of density
profiles is not needed (see \cite{aqua-cornu-profiles1}
\cite{aqua-cornu-profiles2} for a thorough study of density profiles near
boundaries). We only have to introduce weak and plausible assumptions on
the convergence of the profiles to their bulk value.

It is shown in section 4 that the asymptotic value (\ref{0.1a}) of the
force is already obtained at the level of the Debye-H\"uckel theory. The
main tool is the explicit form of the mean-field potential for
piecewise-flat density profiles, related to the potential for structured
profiles by an integral equation. The latter equation is shown to have a
convergent perturbative solution in the weak-coupling regime. At large slab
separation, the Debye-H\"uckel potential factorizes into potentials
pertaining to individual plasmas obeying electroneutrality sum rules.

We establish in section 5 that the theory beyond mean-field does not
provide any additional contribution to the asymptotics (\ref{0.1a}). This
is first done non-perturbatively with the help of integral equations
corresponding to an appropriate dressing transformation of the Ursell
function and under the mild assumption of integrable clustering.  Finally
the result is recovered once again by selecting and resumming the
contribution of dominant Mayer graphs to the full charge correlation at
large separation.

We also treat in section 6 a variant situation where one of the slabs is
replaced by an ideal macroscopic dielectric medium at distance $d$ (namely
generating images of the plasma's fluctuating charges). Using the Green
function of the Poisson equation with appropriate dielectric boundary
conditions, the Lifshitz result for the mean force is retrieved and reduces
to (\ref{0.1a}) as the dielectric constant tends to infinity.  It is
interesting to observe that it is sufficient for the fluctuations to occur
only in one of the bodies to generate the same asymptotic behaviour.

We will come back to the inclusion of magnetic forces and to quantum models
in future works.

\section{Description of the model}

We consider two plasmas $A$ and $B$ of classical point charges confined to
two planar slabs $\LA(L, a)$ and $\LB(L, b)$ in three-dimensional
space. The slabs have thickness $a$ and $b$, surface $L^{2}$, and are
separated by a distance $d$~:
\begin{align}
    &\LA(L, a) := \{\r=(x,\y)\ |\ x \in [-a,0],\ \y\in
    [-\tfrac{L}{2},\tfrac{L}{2}]^2\}\nonumber\\
    &\LB(L, b) := \{\r=(x,\y)\ |\ x \in [d,
        d+b],\ \y\in[-\tfrac{L}{2},\tfrac{L}{2}]^2\} .
\label{1.1}
\end{align}
The plasma $A$ ($B$) is made of charges $e_\alpha$ ($e_\beta$) of species
$\alpha\in S_A$ ($\beta\in S_B$) where $S_{A}$ and $S_{B}$ are index sets
for the species in $\LA(L, a)$ and $\LB(L, b)$ respectively.  We assume
both plasmas to be globally neutral, i.e. carrying no net charge,
\begin{align}
    \sum_{a}e_{\alpha_{a}}=\sum_{b}e_{\beta_{b}}=0
\label{1.2a}
\end{align}
where $\sum_{a}$ ($\sum_{b}$) extends on all particles in $\LA(L, a)$
($\LB(L,b)$). For a particle located at $\r$ we will use the generic
notation $(\gamma,\r)$ where $\gamma\in S_{A} $ if $\r\in \LA(L, a)$ and
$\gamma\in S_{B} $ if $\r\in \LB(L, b)$.  The space external to the slabs
is supposed to have no electrical properties, its dielectric constant being
taken equal to that of vacuum. The charges are confined in the slabs by
hard walls that merely limit the available configuration space to the
regions (\ref{1.1}).

All particles interact via the two-body  potential
\begin{align}
    V(\gamma,\gamma',|\r-\r'|)=e_\gamma e_{\gamma'}v(\r-\r')+\vSR
    (\gamma,\gamma',|\r-\r'|),
\label{1.2}
\end{align}
where $v(\r-\r')= 1/|\r-\r'|$ is the Coulomb potential (in Gaussian units)
and $\vSR(\gamma,\gamma', \r-\r')$ is a short-range repulsive potential to
prevent the collapse of opposite charges and guarantee the thermodynamic
stability of the system.

The total potential energy $U$ consists in the sum of all pairwise
interactions, separated into three contributions according to whether they
take place between two particles of $A$, of $B$, or between a particle of
$A$ and a particle of $B$~:
\begin{align}
    U = U_A + U_B + U_{AB}.
\label{1.3}
\end{align}

On the microscopic level, the force between configurations of charges in
the two plasmas is the sum of all pairwise forces exerted by the particles
of $B$ on the particles of $A$~:
\begin{align}
    &\B{F}_{\LB\to\LA} := \sum_a \sum_b\left[ e_{\alpha_a} e_{\beta_b}
        \frac{\r_a-\r_b}{|\r_a-\r_b|^3} + {\bf F}_{\text{SR}}(\alpha_a,
        \beta_b, \r_a-\r_b)\right]\nonumber\\
    &\r_{a}\in \LA(L, a),\quad \r_{b}\in\LB(L, b)
\label{1.4}
\end{align}
and ${\bf F}_{\text{SR}}$ is the force associated to the short-range
potential $\vSR$.  For simplicity we assume that the range of $\vSR$ is
finite so that ${\bf F}_{\text{SR}}(\alpha_a, \beta_b, \r_a-\r_b)$ vanishes
as soon as $d$ is large enough, and we will omit it in the following.

Both plasmas are supposed to be in thermal equilibrium at the same
temperature T.  The statistical average $\langle\cdots\rangle_{L}$ is
defined in terms of the Gibbs weight $\exp(-\beta U),\,
\beta=(k_{B}T)^{-1}$, associated with the total energy (\ref{1.3}). There
is no need to explicitly specify the ensemble used here (canonical or grand
canonical) provided that the global neutrality constraint (\ref{1.2a}) is
taken into account.  The average particle densities $\rho_L(\gamma\,\r)$
are expressed as averages of the microscopic particle densities
$\hat{\rho}(\gamma\,\r) = \sum_i\delta_{\gamma\,\gamma_i}\delta(\r-\r_i)$
where the sum runs over all particles
\begin{align}
    \rho_L(\gamma\,\r)=\langle \hat{\rho}(\gamma\,\r)\rangle_{L}.
\label{1.5}
\end{align}
We keep the index $L$ to remember that averages are taken for the
finite-volume slabs (\ref{1.1}). Hence expressing the sums in (\ref{1.4})
as integrals on particle densities $\hat{\rho}(\gamma\,\r)$, the average
force reads
\begin{align}
    &\avg{\B{F}}_L = \Int[\LA(L)]{\d\r} \Int[\LB(L)]{\d\r'}
    \frac{\r-\r'}{|\r-\r'|^3}\ c_L(\r, \r')
\label{1.6}
\intertext{where $c_L(\r, \r')$ is the two-point charge correlation
    function}
    &c_L(\r,\r') = \langle\hat{c}(\r)\hat{c}(\r')\rangle_L= \sum_{\gamma,
    \gamma'} e_\gamma e_{\gamma'} \rho_L(\gamma\,\r,\gamma'\, \r'),
	\;\;\;\hat{c}(\r)=\sum_{\gamma} e_\gamma \hat{\rho}(\gamma\,\,\r).
\label{1.6a}
\end{align}

We now consider the average force by unit surface between two infinitely
extended slabs at distance $d$ by letting their transverse dimension $L$
tend to infinity. We assume that the plasma phases are in fluid states
homogeneous and isotropic in the $\y$ directions, namely the charge
correlation has an infinite-volume limit of the form
\begin{equation}
    \lim_{L\to\infty}c_L(\r,\r')=\langle\hat{c}(\r)\hat{c}(\r')\rangle =
    c(x,x',|\y-\y'|).
\label{1.7}
\end{equation}
For symmetry reasons, $\avg{\B{F}}_L $ has no transverse component and is
directed along the $x$ axis perpendicular to the plates.  We therefore
consider the $x$-component of the force per unit surface
\begin{align}
    \avg{f} &:= \lim_{L\to\infty}\frac{\langle F_{x}\rangle_L}{L^{2}}
    =\lim_{L\to\infty} \frac{1}{L^2} \Int[L^{2}]{\d\y}
    \left(\Integral{-a}{0}{\d x}\Integral{d}{d+b}{\d x'}\Int[L^{2}]{\d\y'}
    \frac{x-x'}{|\r-\r'|^3}\, c_L(x,\y,x',\y')\right)\nonumber\\ &=
    \Integral{-a}{0}{\d x}\Integral{d}{d+b}{\d x'}\Int{\d \y}
    \frac{x-x'}{\big[(x-x')^2+|\y|^2\big]^{3/2}}\,
    c(x,x',|\y|).\label{eq:casimirforce2}
\end{align}
The last line results from the $\y$ translational invariance of the
integrand in the limit $L\to\infty$. We do not justify the existence of the
limit here (which depends on an uniform control of $c_L(x,\y, x',\y')$ as
$|\y -\y'|\to\infty$), but it will be clear from the subsequent
calculations that (\ref{eq:casimirforce2}) is a well defined quantity, at
least in the weak-coupling regime.

Formula (\ref{eq:casimirforce2}) remains valid if one replaces
$c(x,x',|\y|)$ by the truncated charge-charge correlation function
\begin{align}
    S(x,x',\y)=\avg{\hat{c}(\r)\hat{c}(\r')} -
    \avg{\hat{c}(\r)}\avg{\hat{c}(\r')} ,\quad \r=(x,\y),\;\,\,\r'=(x',\0)
\label{1.8}
\end{align}
with $\hat{c}(\r)$ the microscopic charge density as in (\ref{1.6a}).
Indeed, the $\y$-Fourier transform of the Coulomb force reads
\begin{align}
    \Int{\d\y} \e^{-i\k\cdot\y}
    \frac{x-x'}{[(x-x')^{2}+|\y|^{2}]^{3/2}}=2\pi \;
    \mathrm{sign}(x-x')\e^{-k|x-x'|}
\label{1.9}
\end{align}
and reduces to $-2\pi$ when $\k =\0$ and $x<x'$. This implies that the
charge density profile $\avg{\hat{c}(\r)}=c(x)$ does not contribute to the
force because of the global neutrality of both plasmas
\begin{align}
    \Integral{-a}{0}{\d x} c(x) = \Integral{d}{d+b}{\d x} c(x) = 0.
\label{1.10}
\end{align}

To take full advantage of the translational invariance in the $\y$
direction we represent the $\y$-integral in (\ref{eq:casimirforce2}) in
Fourier space~:
\begin{align}
    \avg{f}=-\frac{1}{2\pi}\Integral{-a}{0}{\d x}\Integral{d}{d+b}{\d x'}
    \!\!  \Int{\d\k} \e^{-k|x-x'|}S(x,x',\k)
\label{1.11}
\end{align}
where $k=|\k|$ and $S(x,x',\k)=\int\! \d\k\, e^{-i\k\cdot\y} S(x,x',\y)$.
The dependence of $\avg{f}=\avg{f}(d)$ on the separation $d$ between the
two slabs occurs in the integration limits in (\ref{1.11}) as well as in
the charge correlation function $S(x,x',\k)$.  The $d$ dependence of the
correlations between the two slabs $A$ and $B$ originates itself from the
Coulomb interaction term $U_{AB}$ occurring in the total Gibbs thermal
weight. The object of the next sections is to determine the asymptotic
behaviour of $\avg{f}(d)$ as $d\to\infty.$

\section{Mayer series for inhomogeneous charged fluids}\label{sec:mayer}

We briefly summarise the methods that we use to calculate the charge-charge
correlation function of our system. Let us consider a general charged fluid
in presence of spatial inhomogeneities caused by an external potential
$\Psi^{{\rm ext}}(\gamma\,\r)$, e.g. wall potentials confining the system
in some region of space.  Hard walls without electrical properties
(infinite potentials) can be implemented by simply declaring that the
density vanishes in the forbidden regions.

It is well-known (e.g. \cite{hansen-mcdonald}) that the two point Ursell
function, related to the densities $\rho(i),\rho(j)$ and the two-particle
correlation $\rho(i,j)$
\begin{align}
    h(i,j) := \frac{\rho(i,j)}{\rho(i)\rho(j)}-1,
\label{2.1}
\end{align}
can be expanded in a formal power series of the densities by means of Mayer
graphs.  The basic Mayer bonds are
\begin{align}
    f(i,j)= e^{-\beta V(i,j)}-1
\label{2.1a}
\end{align}
where $V(i,j)$ is the potential (\ref{1.2}) and the weights at vertices are
the densities $\rho(i)$. Here $i$ is a shorthand notation for the point
$(\gamma_i\,\r_i)$ in configuration space, and integration on
configurations $\sum_{\gamma_{i}}\int d\r_i$ includes the summation on
particle species. Diagrams have two root points $i$ and $j$ and $m$
internal points which have to be integrated over. Each pair of points is
linked by at most one $f$-bond and there are no articulation
points \footnote{An articulation point, when removed, splits the diagram
    into two pieces, at least one of which is disconnected from the root
    points}.  Because of the long-range of Coulomb interaction, the
integrals occurring in every diagram diverge in the thermodynamic limit. It
is therefore necessary to introduce the screened mean-field potential
$\Phi(\r_{i},\r_{j})$ as usual by resumming the chains built with pure
Coulombic interaction bonds $-\beta e_{\gamma_{i}}
e_{\gamma_{j}}v(\r_{i}-\r_{j})$. Then replacing the bare Coulomb potential
by the screened potential leads to a reorganisation of the diagrammatic
expansion of the Ursell function resulting in the formula
\cite{aqua-cornu-mayer}
\begin{align}
    h(\gamma\,\r,\gamma'\,\r') = \sum_{\Pi} \frac{1}{S_\Pi}
    \sum_{\gamma_1,...,\gamma_m}\Int{\d\r_1\cdots\d\r_m}
    \rho(\gamma_1\,\r_1)\cdots\rho(\gamma_m\,\r_m) \prod_{\{i,j\}\in\Pi}
        {\cal F}(i,j)
\label{eq:hresummeddiagrams}
\end{align}
The first sum runs over all unlabelled topologically different connected
diagrams $\Pi$ (called prototype graphs) with two root points
$(\gamma\,\r)$ and $(\gamma'\,\r')$ and $m$ integrated internal points with
density weights ($m$ ranges from $0$ to $\infty$); $S_\Pi$ denotes the
symmetry number of a diagram $\Pi$.  Each pair of points is linked by at
most one bond ${\cal F}\in\{F,\Fr\}$ and there are no articulation
points. Moreover, convolutions of $F$ bonds are forbidden to avoid multiple
counting of original Mayer graphs.  The two possible bonds read
\begin{align}
    &F(i,j) = -\beta e_{\gamma_i} e_{\gamma_j}
    \Phi(\r_i,\r_j) \label{eq:Fcc}\\
    & \Fr(i,j) = \exp[{-\beta e_{\gamma_i} e_{\gamma_j}
            \Phi(\r_i,\r_j)-\beta \vSR(\gamma_i,\gamma_j,|\r_i-\r_j|)}] -1
    +\beta e_{\gamma_i} e_{\gamma_j} \Phi(\r_i,\r_j).
\label{eq:Fr}
\end{align}
These bonds are obtained in terms of the Debye-H\"uckel screened potential
$\Phi$, which is symmetric and defined as the solution of the integral
equation
\begin{align}
    &\Phi(\r,\r') = v(\r-\r') - \frac{1}{4\pi}\Int{\d\r_1}
    \kappa^2(\r_1)\ v(\r-\r_1)\ \Phi(\r_1,\r')=\Phi(\r',\r)
\label{eq:mfpotential-integraleq}
\end{align}
or equivalently of the differential equation
\begin{align}
    \Delta\Phi(\r,\r')-\kappa^2(\r)\Phi(\r,\r')=-4\pi\delta(\r-\r')
\label{2.2}
\end{align}
supplemented by suitable boundary conditions.  In
(\ref{eq:mfpotential-integraleq}) and (\ref{2.2})
\begin{align}
    \kappa(\r) := \left(4\pi\beta\sum_\gamma
    e_\gamma^2\rho(\gamma\,\r)\right)^{1/2}
\label{eq:kappa}
\end{align}
can be interpreted as the local inverse Debye screening length in the
inhomogeneous system. The bond $\Fr(i,j)$ includes the short-range
contribution and the nonlinear Coulombic part of the original Mayer bond.

The densities $\rho(\gamma\,\r)$ entering in (\ref{eq:hresummeddiagrams})
and (\ref{eq:kappa}) have to be determined self-consistently from the first
equation of the Born-Green-Yvon hierarchy which links the one-point and the
two-point functions.  For charged systems, it takes the form \cite{martin}
\begin{align}
    \nabla\rho(\gamma\,\r)=&-\beta e_{\gamma}\rho(\gamma\,\r)\left[\nabla
        \Psi(\gamma\,\r)+\Int{\d\r'}
        \left(\sum_{\gamma'}e_{\gamma'}\rho(\gamma'\,\r')
        h(\gamma\,\r,\gamma'\,\r')\right)\nabla v(\r-\r')\right]\nonumber\\
    &-\beta\sum_{\gamma'}\Int{\d\r'}
    \rho(\gamma\,\r,\gamma'\,\r')\nabla\vSR(\gamma,\gamma', |\r-\r'|)
\label{2.2a}
\end{align}
where
\begin{align}
    \Psi(\gamma\,\r)=\Psi^{{\rm ext}}(\gamma\,\r)+ \Int{\d\r'} c(\r')
    v(\r-\r')
\label{2.3}
\end{align}
is the sum of the external potential and the electrostatic potential caused
by the inhomogeneous mean charge density $c(\r')$ in the system.  Hence the
Ursell function (considered as a functional of the densities through its
Mayer expansion (\ref{eq:hresummeddiagrams})) together with (\ref{2.2a})
form a closed set of equations whose solution determines in principle the
exact densities and two-particle correlations.  The differential equation
(\ref{2.2a}) has still to be supplemented with appropriate boundary
conditions.  For instance if the system is asymptotically uniform in some
directions, one can fix the corresponding asymptotic bulk densities.

Finally, the charge-charge correlation function (\ref{1.8}) is related to
the Ursell function by
\begin{align}
    S(\r,\r') = \sum_{\gamma,\gamma'} e_\gamma e_{\gamma'}
    \rho(\gamma\,\r)\rho(\gamma'\,\r')h(\gamma\,\r,\gamma'\,\r') +
    \delta(\r - \r')\sum_{\gamma}e_{\gamma}^{2}\rho(\gamma\,\r)
\label{2.4}
\end{align}
The second term in the r.h.s. of (\ref{2.4}) is the contribution of
coincident points.

We shall use the above formalism to calculate (\ref{2.4}) as a function of
the distance $d$ for two infinitely-thick plasmas (i.e $a\to\infty$ and
$b\to \infty$ in (\ref{1.11})) and with hard walls at $x=0$ and $x=d$. In
this situation we take $\Psi^{{\rm ext}}(\gamma\,\r)=0$ for $x<0$ and
$x>d$, and impose
\begin{align}
    \rho(\gamma\,x)&=0, \quad 0\leq x \leq d, \nonumber\\
    \lim_{x\to-\infty}\rho(\gamma\, x)&=\rho_{A\,\gamma},\quad
    \lim_{x\to\infty}\rho(\gamma\, x)=\rho_{B\,\gamma}
\label{2.5}
\end{align}
where $\rho_{A\,\gamma}$ and $\rho_{B\,\gamma}$ are the bulk particle
densities of plasmas $A$ and $B$. The contribution of coincident points
does not enter into the force (\ref{1.11}) since $\r$ and $\r'$ are always
at least separated by the distance $d$.  Therefore, (\ref{1.11}) reads
\begin{align}
    \avg{f}=-\Integral{-\infty}{0}{\d x}\Integral{d}{\infty}{\d x'}\!\!
    \Integral{0}{\infty}{\d k} k e^{-k|x-x'|}\sum_{\gamma,\gamma'} e_\gamma
    e_{\gamma'} \rho(\gamma\,x)\rho(\gamma'\,x')h(\gamma\, x,\gamma'\, x',
    \k)
\label{2.4a}
\end{align}
with $h(\gamma\, x,\gamma'\, x', \k)$ the $\y$-Fourier transform of the
Ursell function.

\section{Debye-H\"uckel theory}\label{sec:fcc}

In this section we show that the simplest contribution to
$h(\gamma\,\r,\gamma'\,\r')$ given by the sole bond $F$, namely,
\begin{align}
    h^{{\rm DH}}(\gamma\,\r,\gamma'\,\r')=-\beta
    e_{\gamma}e_{\gamma'}\Phi(\r,\r')
\label{3.1}
\end{align}
already leads to the asymptotic value (\ref{0.1a}) of the force.  For this
we have to find the screened potential by solving (\ref{2.2}) (written in
Fourier form) with the boundary conditions imposed by the slab geometries
\begin{align}
    \left[\frac{\p^2}{{\p x}^2} - k^2 - \kappa^2(x)\right] \Phi(x,x',\k) =
    -4\pi\delta(x-x'),\quad \kappa(x)=0,\;\;0<x<d,
\label{eq:mfpotential-diffeq}
\end{align}
with $\Phi(x,x',\k)=\int\! \d\y\, \e^{-i\k\cdot\y}\Phi(x,x',\y)$.  The
boundary conditions are $\Phi(x,x',\k)$ and $\p \Phi(x,x',\k)/ \p x$
continuous at $x=0$ and $x=d$, and $\lim_{x\to\pm\infty}\Phi(x,x',\k) =0$.
The density profiles entering in $\kappa^{2}(x)$ by (\ref{eq:kappa}) are
not known (since they have to be determined by self-consistency from
(\ref{2.2a})), but we will not need their explicit form in the sequel
\footnote{A mean-field approximation to the densities could be obtained by
    replacing $h$ in (\ref{2.2a}) by $h^{{\rm DH}}$. We are not doing so
    here but deal throughout with the exact densities.}. We only need to
assume that their difference to bulk value is integrable~:
\begin{align}
    &\rho(\gamma\,x)-\rho_{A\,\gamma} = \mathcal{O}
    \left(\frac{1}{|x|^{1+\epsilon} }\right),\,\, x\to -\infty, \nonumber\\
    &\rho(\gamma\, x) - \rho_{B\,\gamma} =
    \mathcal{O}\left(\frac{1}{|x|^{1+\epsilon}}\right),\,\, x\to \infty,
    \qquad \epsilon > 0.
\label{3.2}
\end{align}
Integrating (\ref{eq:mfpotential-diffeq}) on $x$ leads to
\begin{align}
    \Integral{-\infty}{\infty}{\d x} \frac{\kappa^2(x)}{4\pi} \Phi(x,x',\k)
    = 1 - \frac{k^2}{4\pi}\Int{\d x} \Phi(x,x',\k).
\label{3.3a}
\end{align}
In particular, for $\k=0$
\begin{align}
    \Integral{-\infty}{\infty}{\d x}
    \frac{\kappa^{2}(x)}{4\pi}\Phi(x,x',\k=0)=1
\label{3.3}
\end{align}
which is nothing else than the electroneutrality sum rule for the
charge-charge correlation (\ref{1.8}), (\ref{2.4}) within the Debye regime
(\ref{3.1})~ \cite{martin}:
\begin{align}
    \Integral{-\infty}{\infty}{\d x} \Int{\d\y} S^{DH}(x,x',\y) = 0
\label{3.4}
\end{align}

To solve (\ref{eq:mfpotential-diffeq}) we first consider the simpler
problem with piecewise-flat densities $\rho_{A\,\gamma}$ and
$\rho_{B\,\gamma}$ in each plasma
\begin{align}
    &\left[\frac{\p^2}{{\p x}^2} - k^2 - \bar{\kappa}^2(x)\right]
    \varphi(x,x',\k) = -4\pi\delta(x-x')\label{3.5}\\
    & \bar{\kappa}(x)=\kappa_{A},\;x<0,\quad \bar{\kappa}(x)=0, \;0<x<d,
    \quad\bar{\kappa}(x) = \kappa_{B},\;x>d\nonumber
\end{align}
where
\begin{align}
    \kappa_{A}=\left(4\pi\beta \sum_{\alpha \in
        S_{A}}e^{2}_{\alpha}\;\rho_{A\,\alpha}\right)^{1/2},\quad
    \kappa_{B}=\left(4\pi\beta \sum_{\beta \in
        S_{B}}e^{2}_{\beta}\;\rho_{B\,\beta}\right)^{1/2}
\end{align}
are the bulk inverse screening lengths.  The boundary conditions are the
same as for (\ref{eq:mfpotential-diffeq}).  Denoting ${\mathcal L}$ the
linear operator acting on $\Phi$ on the l.h.s. of
(\ref{eq:mfpotential-diffeq}) and by $\bar{{\mathcal L}}$ the one acting
similarly on $\varphi$, one has ${\mathcal L}\Phi(x)-\bar{{\mathcal
        L}}\varphi(x) = \bar{{\mathcal L}} (\Phi-\varphi)(x)-u(x)\Phi(x)
=0$, where $u(x) = \kappa^2(x)-\bar{\kappa}^2(x)$ represents the deviation
of the density profiles to their bulk limiting values. Since
$-\varphi/4\pi$ is the Green function of $\bar{{\mathcal L}}$, it follows
that $\Phi(x,x',\k)$ and $\varphi(x,x',\k)$ are related by the integral
equation
\begin{align}
    \Phi(x,x',\k) = \varphi(x,x',\k) - \frac{1}{4\pi} \Int{\d s}
    u(s)\ \varphi(x,s,\k)\ \Phi(s,x',\k)
    \label{eq:mfpotential-integraleq2}
\end{align}
which expresses $\Phi(x,x',\k)$ as a perturbation of $\varphi(x,x',\k)$ by
the inhomogeneity $u(x)$ of the plasmas' density profiles.

Solving (\ref{3.5}) piecewise and connecting the solutions together yields
\footnote{The functions $\varphi_{AB},\;\varphi_{BB}$ (depending on $d$)
    refer to the system of the two plasmas under mutual influence with the
    $x$-location of particles in plasma $B$ measured by their distance from
    the boundary at $d$ (i.e. from $0$ to $+\infty$). In the sequel, the
    quantities $\varphi_{A}^{0}$, $\varphi_{B}^{0}$ (independent of $d$)
    refer similarly to the single semi-infinite plasma $A$ and
    $B$.\label{foot:B-shift}}
\begin{align}
    &\varphi(x,x',\k) = \begin{cases}
        \varphi_{AA}(x,x',\k),&\quad x,x'<0 ,\\
        \varphi_{AB}(x,x'-d,\k), &\quad x<0<d<x',\\
        \varphi_{BB}(x-d,x'-d,\k),&\quad d<x,x',
        \end{cases}
\label{eq:mfflatpotential}
\intertext{with}
	&\varphi_{AA}(x,x',\k) = 2\pi \frac{\e^{-k_A|x-x'|}}{k_A} +
	2\pi\frac{\e^{-k_A|x+x'|}}{k_A} \tfrac{(k_A-k)(k_B+k)
        \e^{kd}-(k_A+k)(k_B-k) \e^{-kd}}{(k_A+k)(k_B+k) \e^{kd}-(k_A-k)
        (k_B-k)\e^{-kd}},
\label{3.5a}\\
	&\varphi_{AB}(x,x',\k) = \frac{8\pi k \e^{-k_A|x|}
    \e^{-k_B|x'|}}{(k_A+k)(k_B+k)\e^{kd} - (k_A-k)(k_B-k)\e^{-kd}},
\label{3.5b}\\
    &k_{A} = \sqrt{k^2+\kappa_{A}^2},\quad k_{B} = \sqrt{k^2+\kappa_{B}^2}.
\end{align}
The function $\varphi_{BB}(x,x',\k)$ is obtained by interchanging the
indices $A$ and $B$ in (\ref{3.5a}). Notice that
$\varphi(x,x',\k)=\varphi(x',x,\k)$ and is invariant under the symmetry
$x\leftrightarrow d-x, x'\leftrightarrow d-x', A\leftrightarrow B$.

We discuss a few properties of this solution.  The first term in the r.h.s
of (\ref{3.5a}) corresponds to the bulk Debye-H\"uckel potential whereas
the second term is the modification due to the finite boundaries of both
plasmas $A$ and $B$.  As $d\to \infty$, $\varphi_{AA}(x,x',\k)$ reduces to
the well-known Debye-H\"uckel potential $\varphi_{A}^{0}(x,x',\k)$ of a
single semi-infinite plasma in the region $x<0$ (see formula
(24) in \cite{phiA0} and \cite{jancovici-along-the-wall})
\begin{align}
    \lim_{d\to\infty}\varphi_{AA}(x,x',\k)= 2\pi
    \frac{\e^{-k_A|x-x'|}}{k_A}+2\pi\frac{k_A-k}{k_A+k}
    \frac{\e^{-k_A|x+x'|}}{k_{A}}=\varphi_{A}^{0}(x,x',\k)
\label{3.6}
\end{align}
uniformly with respect to $\k$. One observes that $\varphi(x,x',\k)$ is an
even, infinitely differentiable function of $|\k|$, implying that
$\varphi(x,x',\y)$ decays along walls directions faster than any inverse
power of $y$.  This is to be contrasted with the small $\k$ behaviour of
the function (\ref{3.6}) which has a non analytic $|\k|$ term leading to
the algebraic decay $y^{-3}$ along the wall
\cite{jancovici-along-the-wall}; see also \cite{martin}, sect. III.C.2.

The function $\varphi_{AB}(x,x',\k)$ (\ref{3.5b}) describes the correlation
between the two plasmas. In terms of the scaled dimensionless variable
$\q=\k\,d$, it has the simple factorized asymptotic behaviour
\begin{align}
    \varphi_{AB}(x,x',\tfrac{\q}{d}) &\sim \frac{1}{d}\frac{4\pi
        q}{\kappa_{A}\kappa_{B}\sinh q}\,
    \e^{-\kappa_A|x|}\,\e^{-\kappa_B|x'|} \nonumber\\
    &=\frac{1}{d}\frac{q}{4\pi\sinh q} \,\varphi_{A}^{0}(x,0,\0)
    \,\varphi_{B}^{0}(0,x',\0),\quad d\to\infty
\label{3.7}
\end{align}

Finally $\varphi(x,x',\k)$ obeys the following bound uniformly with respect
to $\k$ and $d$ (appendix B)
\begin{align}
    &0 \leq \varphi(x,x',\k) \leq \varphi^>(x,x') \leq
    \frac{4\pi}{\kappa},\quad \kappa d \geq 1
\label{eq:mfflatpotential-bounds}
\end{align}
The function $\varphi^>(x,x')$ is defined as in (\ref{eq:mfflatpotential})
with $\varphi_{AA}$, $\varphi_{AB}$ and $\varphi_{BB}$ replaced by
\begin{align}
    &\varphi_{AA}^>(x,x') \Doteq \varphi_{BB}^>(x,x') \Doteq
    \frac{2\pi}{\kappa}\left(\e^{-\kappa|x-x'|} + \e^{-\kappa|x+x'|}\right)
\label{3.8}\\
     &\varphi_{AB}^>(x,x') \Doteq \varphi_{BA}^>(x,x') \Doteq
	\frac{4\pi}{\kappa^2 d}
    \e^{-\kappa|x|}\e^{-\kappa|x'|},\quad \kappa := \min\{\KA,\KB\}
\label{3.9}
\end{align}

The Debye-H\"uckel potential $\Phi(x,x',\k)$ can be obtained by iterating
the integral equation (\ref{eq:mfpotential-integraleq2}). Convergence can
be established in the weak-coupling regime~:

\vspace{2mm}

\noindent {\bf Lemma} (see proof in appendix B)

\vspace{2mm}
{\it
\noindent Let
\begin{align}
    r := \frac{1}{\kappa}\Int{\d x |u(x)|}= \frac{1}{\kappa}\Int{\d x
        |\kappa^{2}(x)-\bar{\kappa}^{2}(x)| }
\label{3.10}
\end{align}
Then for $r<1$ (\ref{eq:mfpotential-integraleq2}) has a solution with the
bound
\begin{align}
    \left|\Phi(x,x',\k)\right| \leq
    \frac{1}{1-r}\varphi^>(x,x').\label{eq:mfpotential-bounds}
\end{align}
}

As in (\ref{eq:mfflatpotential}) we distinguish various contributions
according to the location of the arguments $x,x'$ of $\Phi(x,x',\k)$ by
setting (see footnote \ref{foot:B-shift})
\begin{align}
    &\Phi(x,x',\k) = \begin{cases}\Phi_{AA}(x,x',\k)\ ,&\quad x,x'<0,\\
        \Phi_{AB}(x,x'-d,\k)\ , &\quad x<0<d<x',\\
        \Phi_{BB}(x-d,x'-d,\k)\ ,&\quad d<x,x',
    \end{cases}
\label{3.12}
\end{align}
\begin{align}
    &\rho(\gamma\, x)= \begin{cases} \rho_{A}(\gamma\, x)\ , &\quad x<0,\\
        \rho_{B}(\gamma, x-d)\ , &\quad x>d.\end{cases}
\label{3.12a}
\end{align}
The quantities $\kappa_{A,B}(x),\,u_{A,B}(x)$ are defined in the same way.
Then the integral equation (\ref{eq:mfpotential-integraleq2}) splits into
\begin{align}
    \Phi_{AA}(x,x',\k) = \varphi_{AA}(x,x',\k) &-
    \frac{1}{4\pi}\Integral{-\infty}{0}{\d s} u_A(s) \varphi_{AA}(x,s,\k)
    \Phi_{AA}(s,x',\k)\nonumber\\
    &- \frac{1}{4\pi}\Integral{0}{\infty}{\d s} u_B(s) \varphi_{AB}(x,s,\k)
    \Phi_{BA}(s,x',\k).
\label{3.13}\\
    \Phi_{AB}(x,x',\k) = \varphi_{AB}(x,x',\k) &-
    \frac{1}{4\pi}\Integral{-\infty}{0}{\d s} u_A(s) \varphi_{AA}(x,s,\k)
    \Phi_{AB}(s,x',\k)\nonumber\\
    &- \frac{1}{4\pi}\Integral{0}{\infty}{\d s} u_B(s) \varphi_{AB}(x,s,\k)
    \Phi_{BB}(s,x',\k).
\label{3.14}
\end{align}

The density profiles depend on $d$ because of the mutual Coulomb
interactions between the two plasmas.  We shall examine the asymptotic
behaviour of $\Phi(x,x',\k)$ as $d\to\infty$ under the assumption that
these density profiles are uniformly bounded with respect to $d$ and tend
to those of single semi-infinite plasmas i.e.
\begin{align}
    \lim_{d\to\infty}\rho_{A}(\gamma\,x)=\rho_{A}^{0}(\gamma\,x), \,x<0,
    \quad
    \lim_{d\to\infty}\rho_{B}(\gamma\,x)=\rho_{B}^{0}(\gamma\,x), \, x>0
\label{limprof}
\end{align}
We denote by $\kappa_{A,B}^{0}(x),\,u_{A,B}^{0}(x)$ the analogous
quantities for the single semi-infinite plasmas. Then one concludes from
(\ref{3.13}) that uniformly in $\k$
\begin{align}
    \lim_{d\to\infty}\Phi_{AA}(x,x',\k)=\Phi_{A}^{0}(x,x',\k),\quad x,x'<0
\label{3.14a}
\end{align}
where $\Phi_{A}^{0}(x,x',\k)$ is the Debye-H\"uckel potential of a
semi-infinite plasma in the region $x<0$ determined in terms of the
corresponding flat profile potential $\varphi_{A}^{0}(x,x',\k)$ (\ref{3.6})
by
\begin{align}
    \Phi_{A}^{0}(x,x',\k)=\varphi_{A}^{0}(x,x',\k)-\frac{1}{4\pi}
    \Integral{-\infty}{0}{\d s} u_A^{0}(s) \varphi_{A}^{0}(x,s,\k)
    \Phi_{A}^{0}(s,x',\k),\;\,x,x'<0
\label{3.15}
\end{align}
Indeed, in view of the limits (\ref{3.6}), (\ref{limprof}) and using
dominated convergence with the bounds (\ref{eq:mfflatpotential-bounds}),
(\ref{eq:mfpotential-bounds}) the integral equation (\ref{3.13}) reduces to
(\ref{3.15}) in the limit $\d\to\infty$.  One has likewise
\begin{align}
    \lim_{d\to\infty}\Phi_{BB}(x,x',\k)=\Phi_{B}^{0}(x,x',\k),\quad x,x'>0
\label{3.14b}
\end{align}
where $\Phi_{B}^{0}(x,x',\k)$ is the Debye-H\"uckel potential of a
semi-infinite plasma in the region $x>0$.

We come now to the correlation $\Phi_{AB}(x,x',\tfrac{\q}{d})$ which is
expected to decay as $d^{-1}$ at large separation of the two plasmas. To
see this it is useful to write (\ref{3.14}) in an alternative form such
that $\varphi_{AB}$ appears explicitly in each term of the equation~:
\begin{align}
    &\Phi_{AB}(x,x',\k) = \varphi_{AB}(x,x',\k)\nonumber \\
    &- \frac{1}{4\pi}\Integral{-\infty}{0}{\d s}
    u_A(s)\,\widetilde{\Phi}_{AA}(x,s,\k)\, \varphi_{AB}(s,x',\k)
    - \frac{1}{4\pi}\Integral{0}{\infty}{\d s}
    u_B(s)\,\varphi_{AB}(x,s,\k)\, \Phi_{BB}(s,x',\k)\nonumber \\
    &+\left(\tfrac{1}{4\pi}\right)^2\!\!\Integral{-\infty}{0}{\d s_1}\!\!
    \Integral{0}{\infty}{\d s_2} u_A(s_1)u_B(s_2)\,
    \widetilde{\Phi}_{AA}(x,s_1,\k) \,\varphi_{AB}(s_1,s_2,\k)
    \,\Phi_{BB}(s_2,x', \k).
\label{3.16}
\end{align}
Here $\widetilde{\Phi}_{AA}(x,x',\k)$ verifies the equation
\begin{align}
	\widetilde{\Phi}_{AA}(x,x',\k) = \varphi_{AA}(x,x',\k) - \frac{1}{4\pi}
    \Integral{-\infty}{0}{\d s} u_A(s)
    \varphi_{AA}(x,s,\k)\ \widetilde{\Phi}_{AA}(s,x',\k),
\label{eq:mfpotential-almost-isol-parts}
\end{align}
which is the equation (\ref{3.13}) with the last term omitted. Equation
(\ref{3.16}) is obtained by iterating (\ref{3.14}) and resumming the
$\varphi_{AA}$ chains to $\widetilde{\Phi}_{AA}$, or by verifying that it
satisfies the basic differential equation (\ref{eq:mfpotential-diffeq}). By
the same arguments that led to (\ref{3.14a}), it is clear that
$\widetilde{\Phi}_{AA}$ also tends to the potential $\Phi_{A}^{0}$ of the
semi-infinite plasma.

Introducing the limits (\ref{3.7}), (\ref{limprof}), (\ref{3.14a}),
(\ref{3.14b}) in (\ref{3.16}) and using again dominated convergence
provided by the bounds (\ref{3.9}), (\ref{3.8}),
(\ref{eq:mfpotential-bounds}) we find that
\begin{align}
    \lim_{d\,\to\infty}d\,\Phi_{AB}(x,x',\tfrac{\q}{d})&=\frac{q}{4\pi\sinh
        q}\nonumber\\
    &\times\left[\varphi_{A}^{0}(x,0,\0) - \frac{1}{4\pi}
        \Integral{-\infty}{0}{\d s} u_A^{0}(s)
        \Phi_{A}^{0}(x,s,\0)\varphi_{A}^{0}(s,0,\0)\right] \nonumber\\
    &\times\left[\varphi_{B}^{0}(0,x',\0) - \frac{1}{4\pi}
        \Integral{0}{\infty}{\d s} u_B^{0}(s)
        \varphi_{B}^{0}(0,s,\0)\Phi_{B}^{0}(s,x',\0)\right] \nonumber\\
    &=\frac{q}{4\pi\sinh q}\Phi_{A}^{0}(x,0,\0)\Phi_{B}^{0}(0,x',\0)
\label{3.17}
\end{align}
As in (\ref{3.7}), the limit factorizes into the product of Debye-H\"uckel
potentials for single semi-infinite plasmas evaluated with one point on the
boundary.  The last line of (\ref{3.17}) follows from (\ref{3.15}), the
corresponding equation for $\Phi_{B}^{0}(x,x',\0)$, and the fact that these
functions are symmetric in $x,x'$.

With this result we can determine the leading term in the asymptotic
behaviour of the force (\ref{1.11}) in the Debye-H\"{u}ckel regime.  From
(\ref{2.4a}) and (\ref{3.1}), one has
\begin{align}
	&\avg{f}^{\text{DH}}(d)= \frac{1}{\beta}\Integral{-\infty}{0}{\d
        x}\Integral{d}{\infty}{\d x'}\!\!\Integral{0}{\infty}{\d
        k}k\,\e^{-k|x-x'|}\,\frac{\kappa^{2}(x)}{4\pi}
    \frac{\kappa^{2}(x')}{4\pi} \Phi(x,x',\k)\nonumber\\
    &=\frac{1}{\beta d^{2}}\Integral{-\infty}{0}{\d
        x}\Integral{0}{\infty}{\d x'}\!\!  \Integral{0}{\infty}{\d q}
    q\,\e^{-\frac{q}{d}|x-x'+d|}\,
    \frac{\kappa_{A}^{2}(x)}{4\pi}\frac{\kappa_{B}^{2}(x')}{4\pi}
    \Phi_{AB}(x,x',\tfrac{\q}{d})
\label{3.18}
\end{align}
To obtain the second line we have set $\k\, d= \q$, shifted the
$x'$-integration by $-d$, and introduced the notation
(\ref{3.12}),(\ref{3.12a}). As $d\to\infty$, one can use (\ref{3.17}) and
the bounds (\ref{3.8}), (\ref{eq:mfpotential-bounds}) to conclude again by
dominated convergence that
\begin{align}
    \lim_{d\to\infty}d^3\,\avg{f}^{\text{DH}}(d) &= \frac{1}{8\pi\beta}
    \Integral{0}{\infty}{\d q} \frac{4q^2\e^{-q}}{\e^q-\e^{-q}}
    \left(\Integral{-\infty}{0}{\d x}
    \frac{(\kappa_{A}^{0})^2(x)}{4\pi}\Phi_{A}^{0}(x,0,\0)\right)\nonumber\\
    &\times\left(\Integral{0}{\infty}{\d x'}
    \frac{(\kappa_{B}^{0})^2(x')}{4\pi}\Phi_{B}^{0}(0,x',\0)\right)
    =\frac{\zeta(3)}{8\pi\beta}
\label{3.19}
\end{align}
Indeed, because of the charge sum rules (\ref{3.3}) for the semi-infinite
plasmas, the two parenthesis are equal to $1$, whereas the $q$ integral
yields the value $\zeta(3)$ where $\zeta$ is the Riemann $\zeta$-function.

\section{Contributions of the other graphs}

In this section we show that the single Debye-H\"uckel bond $F$ saturates
the asymptotic behaviour of the force i.e. taking into account the full set
of other diagrams does not modify the result (\ref{3.19}). For this we use
the method of ``dressing of the root points'' that has been introduced in
\cite{dressing-of-the-root-points} to analyse the decay of the quantum
truncated charge correlation function.  Having singled out the contribution
of the single resummed bond $F$, all remaining diagrams constituting
$h(\gamma\,\r,\gamma'\,\r')$ can be classified into four types, depending
on whether their root points are linked to the rest of the diagram by a
single bond $F$ or not
\footnote{A point in a prototype diagram which is linked to the rest of the
    diagram by exactly one $F$ bond is called a Coulomb point in
    \cite{dressing-of-the-root-points}.}. We thus write their sum in the
form
\begin{align}
    h^{\text{R}} := h - F = h^{\text{cc}} + h^{\text{cn}} + h^{\text{nc}} +
    h^{\text{nn}},\label{eq:hr}
\end{align}
where $h^{\text{cc}}$ stands for the contribution of all graphs that do
begin and do end with an $F$ bond (with anything in between),
$h^{\text{cn}}$ for the contribution of those that do begin but do not end
with an $F$ link, and so on. The latter quantities are obviously related to
$h^{\text{nn}}$~ by the following integral equations (notations are as in
sect.  {\ref{sec:mayer}})
\begin{align}
    h^{\text{cn}}(a,b) &:= \Int{\d 1} F(a,1)\rho(1)
    h^{\text{nn}}(1,b)\nonumber\\
    h^{\text{cc}}(a,b) &:= \Int{\d 1}\Int{\d 2}
    F(a,1)\rho(1) h^{\text{nn}}(1,2)\rho(2)F(2,b)
\label{4.1}
\end{align}
and analogously for $h^{\text{nc}}$. Using these representations in
(\ref{eq:hr}) together with (\ref{eq:Fcc}) and building the charge-charge
correlation corresponding to $h^{\text{R}}$ according to (\ref{2.4}) yields
\begin{align}
    & S^{\text{R}}(x,x',\k) = \sum_{\gamma_1}\Int{\d
        x_1}\sum_{\gamma_2}\Int{\d x_2} \left(\delta(x-x_1) -
    \frac{\kappa^2(x)}{4\pi}\Phi(x,x_1,\k)\right)\notag\\
    &\times
    e_{\gamma_1}\rho(\gamma_1\,x_1)\
    h^{\text{nn}}(\gamma_1\,x_1,\gamma_2\,x_2,\k)\
    e_{\gamma_2}\rho(\gamma_2\,x_2) \left(\delta(x_2-x') -
    \frac{\kappa^2(x')}{4\pi}\Phi(x_2,x',\k)\right).
\label{4.2}
\end{align}
The function $ h^{\text{nn}}(\gamma_1\,x_1,\gamma_2\,x_2,\k)$ embodies a
resummed contribution, not explicitly known at this point, of higher order
graphs. The only assumption needed on this function in the sequel is
integrable clustering uniformly with respect to $d$
\begin{align}
    \Integral{-\infty}{\infty}{\d x_{1}}
    h^{\text{nn}}(\gamma_1\,x_1,\gamma_2\,x_2,\k)&<\infty \label{4.3}\\
    \Integral{-\infty}{0}{\d x_{1}}\Integral{d}{\infty}{\d x_{2}}
    h^{\text{nn}}(\gamma_1\,x_1,\gamma_2\,x_2,\k)&<\infty
\label{4.4}
\end{align}
As a consequence of the bounds (\ref{eq:mfflatpotential-bounds}),
(\ref{eq:mfpotential-bounds}) the condition (\ref{4.4}) obviously holds for
the Debye potential $\Phi$, and it is expected to hold for the Ursell
function on the ground that as $x_{1}\to -\infty$ ($x_{2}\to \infty$)
$h^{\text{nn}}(\gamma_1\,x_1,\gamma_2\,x_2,\k)$ has a fast decay in the
bulk part of plasma $A$ (plasma $B$).  Note that integrating (\ref{4.2}) on
$x$ (or $x'$) at $\k =0$ proves the validity of the charge sum rule for the
exact charge-correlation function (see (\ref{3.4}))
\begin{align}
    \Integral{-\infty}{\infty}{\d x}\Int{\d\y} S(x,x',\y) = 0, \quad
    S(x,x',\y)=S^{DH}(x,x',\y)+S^{R}(x,x',\y)
\label{4.5}
\end{align}

Proceeding as in (\ref{3.18}) the contribution of $S^{R}(x,x',\y)$ to the
average force can be written in the form (permuting the $x,x'$ and
$x_{1},x_{2}$ integrals)
\begin{align}
    \avg{f}^{\text{R}}(d) &= -\frac{1}{d^2} \Integral{0}{\infty}{\d q} q
    \Int{\d x_1}\Int{\d x_2} H_1(x_1,\tfrac{\q}{d})\nonumber\\
    \times\sum_{\gamma_1,\gamma_2} e_{\gamma_1}e_{\gamma_2}&\rho(\gamma_1\,
    x_1)\rho(\gamma_2\, x_2) h^{\text{nn}}(\gamma_1\,x_1,\gamma_2\,x_2,
    \tfrac{\q}{d},d) H_2(x_2,\tfrac{\q}{d})
\label{4.6}
\end{align}
where
\begin{align}
    H_1(x_1,\tfrac{\q}{d}) &= \Integral{-\infty}{0}{\d x}
    \left(\delta(x-x_1)-\frac{\kappa^2(x)}{4\pi}
    \Phi(x,x_1,\tfrac{\q}{d})\right) \e^{\frac{\q}{d} x}\nonumber\\
    H_2(x_2,\tfrac{\q}{d}) &=\Integral{d}{\infty}{\d x'}
    \left(\delta(x_{2}-x')-\frac{\kappa^2(x')}{4\pi}
    \Phi(x_2,x',\tfrac{\q}{d})\right) \e^{-\frac{\q}{d}x'}
\label{4.7}
\end{align}
The behaviour of $\avg{f}^{\text{R}}(d)$ as $d\to\infty$ is determined by
that of the functions $H_{1}$ and $H_{2}$, because $
h^{\text{nn}}(\gamma_1\,x_1,\gamma_2\,x_2, \frac{\q}{d},d)$ does not vanish
in the limit when the variables $x_1$ and $x_2$ are both located in the
same plasma, but tends to the corresponding functions associated with a
single semi-infinite plasma.  Both $H_{1}$ and $H_{2}$ are
$\mathcal{O}(1/d)$ so that $\avg{f}^{\text{R}}(d)=\mathcal{O}(1/d^4)$ does
not contribute to the asymptotic behaviour of the force. More precisely,
integrating (\ref{eq:mfpotential-diffeq}) on $x$ gives for $x_{1}<0$
\begin{align}
    &\Integral{-\infty}{0}{\d x}
    \left(\delta(x-x_1)-\frac{\kappa^2(x)}{4\pi}\Phi(x,x_1,\k)\right)
    =\Integral{d}{\infty}{\d x} \frac{\kappa^2(x)}{4\pi} \Phi(x,x_1,\k) +
    \frac{k^2}{4\pi}\Int{\d x} \Phi(x,x_1,\k)\nonumber\\
    &=\Integral{0}{\infty}{\d x} \frac{\kappa^2_{B}(x)}{4\pi}
    \Phi_{BA}(x,x_1,\k) + \frac{k^2}{4\pi}\Int{\d x} \Phi(x,x_1,\k)
\label{4.8}
\end{align}
implying with (\ref{3.8}), (\ref{eq:mfpotential-bounds})
\begin{align}
    H_{1}(x_{1},\tfrac{\q}{d}) = \mathcal{O}\left(\frac{1}{d}\right)
    +\mathcal{O} \left(\frac{q^{2}}{d^{2}}\right),\quad x_{1}<0 \label{4.9}
\end{align}
For $x_{1}>d$ one has
\begin{align}
    H_{1}(x_{1},\tfrac{\q}{d})= -\Integral{-\infty}{0}{\d x}\e^{q
        x/d}\ \frac{\kappa^2(x)}{4\pi}\Phi(x,x_1,\tfrac{\q}{d})
    =\mathcal{O}\left(\frac{1}{d}\right)\e^{-\kappa_{B}(x_{1}-d)}
\label{4.10}
\end{align}
In the same way
\begin{align}
    H_{2}(x_{2},\tfrac{\q}{d})&=\mathcal{O}\left(\frac{\e^{-q}}{d}\right)
    \e^{-\kappa_{A}| x_{2}|},\quad x_{2}<0,\nonumber\\
    H_{2}(x_{2},\tfrac{\q}{d})&=\mathcal{O}\left(\frac{\e^{-q}}{d}\right) +
    \mathcal{O} \left(\frac{q^{2}\e^{-q}}{d^{2}}\right), \quad x_{2}>d
\label{4.11}
\end{align}
(the factor $\e^{-q}$ comes from $\e^{-qx/d}\leq \e^{-q}$ for $x\geq d$ in
(\ref{4.7})).  Inserting these estimates in the four integration domains
determined in (\ref{4.6}) by $x_{1},\;x_{2}<0$, $x_{1},\;x_{2}>d$ together
with the integrability assumptions (\ref{4.3}), (\ref{4.4}) on
$h^{\text{nn}}$ leads to the result
\begin{align}
    \lim_{d\to\infty}d^3\avg{f}(d)=\lim_{d\to\infty}
    d^3\avg{f}^{\text{DH}}(d)+ \lim_{d\to\infty}d^3\avg{f}^{\text{R}}(d) =
    \frac{\zeta(3)}{8\pi\beta}
\label{4.12}
\end{align}

To conclude this section we present an alternative derivation of the result
(\ref{4.12}) by selecting the class of diagrams that give the dominant
contribution to the Ursell function as $d\to\infty$.  For this we decompose
the bond $F(\gamma\, x,\gamma'\, x',\k)$ (in Fourier representation) into
the sum of four terms according to the location of the arguments $x,\;x'$
\begin{align}
    &F=F_{AA}+F_{AB}+F_{BA}+F_{BB}\nonumber\\
    &F_{AA}(\gamma\, x,\gamma'\,
    x',\k)=\begin{cases}F(\gamma\,x,\gamma'\,x', \k)\ , &\quad x,x'<0 \\
    0\ ,&\quad \text{otherwise} \nonumber  \end{cases}\nonumber\\
    &F_{AB}(\gamma\,x,\gamma'\,x',\k)=\begin{cases}F(\gamma\,x,
    \gamma',x'+d,\k)\ , &\quad x<0,x'>0 \\
    0\ ,&\quad \text{otherwise}\end{cases}
\label{4.13}
\end{align}
with $F_{BA}$ and $F_{BB}$ defined likewise and the similar decomposition
for $F^{R}$ ($x$-integrals in plasma $B$ from now on run in the interval
$[0,\infty )$, see footnote \ref{foot:B-shift}). The set of prototype
    graphs is then expanded in a larger set of graphs defined in terms of
    these bonds.  It follows from (\ref{3.14a}) that $F_{AA} $ and
    $F_{AA}^{R}$ bonds have limits $F_{A}^{0}$ and $F_{A}^{0\,R}$ as
    $d\to\infty$ where $F_{A}^{0}$ and $F_{A}^{0\,R}$ are the bonds
    corresponding to the semi-infinite plasma $\Lambda_{A}$ alone and
    likewise for $BB$ bonds.

It is shown in appendix C that the dominant part of the Ursell function
$h_{AB}(\gamma_{a}\,\r_{a},\gamma_{b}\,\r_{b})$ as $d\to\infty$ is
constituted be the set of graphs that have exactly one $F_{AB}$ bond. This
class is obtained by linking the extremity $\gamma_{1}\,\r_{1}$ of
$F_{AB}(\gamma_{1}\,\r_{1},\gamma_{2}\,\r_{2})$ to the root point
$\gamma_{a}\,\r_{a}$ of $h_{AB}(\gamma_{a}\,\r_{a},\gamma_{b}\,\r_{b})$ in
plasma $\Lambda_{A}$ by all possible subgraphs comprising only $AA$ bonds
(otherwise one would introduce additional $AB$ bonds), taking into account
the excluded convolution rule for $F$ bonds. In the same way the other
extremity $\gamma_{2}\,\r_{2}$ of
$F_{AB}(\gamma_{1}\,\r_{1},\gamma_{2}\,\r_{2})$ is linked to the root point
$\gamma_{b}\,\r_{b}$ in plasma $\Lambda_{B}$ by all possible subgraphs made
of $BB$ bonds. One finds in this way
\begin{align}
    h_{AB}(a,b,\k)&\sim\Int{\d 1} \Int{\d 2}
    \left[\delta(a,1)+\left(h_{AA}^{nn}(a,1,\k) +
        h_{AA}^{cn}(a,1,\k)\right) \rho_{ A}(1)\right] \nonumber\\
    &\times F_{AB}(1,2,\k)\left[\delta(2,b) + \left(h_{BB}^{nn}(2,b,\k) +
        h_{BB}^{cn}(2,b,\k)\right) \rho_{B}(2)\right]
\label{4.14}
\end{align}
Here $a=(\gamma_{a}\,x_{a})$, $1=(\alpha_{1}\,x_{1})$,
$2=(\beta_{2}\,x_{2})$, $b=(\gamma_b,x_b)$, and the integration $\int d1
=\sum_{\alpha_{1}}\int_{-\infty}^{0}dx_{1}$ runs on plasma $A$ and $\int d2
=\sum_{\beta_{2}}\int^{\infty}_{0}dx_{2}$ on plasma $B$.  We have also used
that the convolution of translation invariant functions in the
$\y$-direction is the product of their Fourier transforms. The functions
$h_{AA}^{cn}$ and $h_{AA}^{nn}$ are defined as in (\ref{eq:hr}) in terms of
$AA$ bonds (similarly for $h_{BB}^{cn}$ and $h_{BB}^{nn}$ in terms of $BB$
bonds). One can write (\ref{3.17}) in the form
\begin{align}
    F_{AB}(1,2,\tfrac{\q}{d})\sim -\frac{q}{4\pi\beta d \sinh
        q}\frac{F_{A}^{0}(1,a_{0},\0)}{e_{\alpha_{0}}}
    \frac{F_{B}^{0}(b_{0},2,\0)}{e_{\beta_{0}}},\quad d\to\infty
\label{4.15}
\end{align}
where $a_{0}=(\alpha_{0}\,0)$ indexes a charge $e_{\alpha_{0}}$ located at
the boundary $x_{a_{0}}=0$ of $\Lambda_{A}$ and $b_{0}$ indexes a charge
$e_{\beta_{0}}$ at the boundary of $\Lambda_{B}$.  Taking also into account
that the functions $h_{AA}^{cn}$ and $h_{AA}^{nn}$ approach the
corresponding values $(h_{A}^{0})^{cn}$ and $(h_{A}^{0})^{nn}$ of a single
semi-infinite plasma, one finds that the leading term $\sim 1/d$ of
$h_{AB}(\gamma\,x, \gamma'\,x',\tfrac{\q}{d})$ factorizes as
\begin{align}
    h_{AB}(a,b,\tfrac{\q}{d})\sim  -\frac{q}{4\pi\beta d \sinh q}
    \frac{G^{0}_{A}(a,a_{0})}{e_{\alpha_{0}}}
    \frac{G^{0}_{B}(b_{0},b)}{e_{\beta_{0}}},\quad d\to\infty
\label{4.16}
\end{align}
with
\begin{align}
    G^{0}_{A}(a,a_{0})&=F^{0}_{A}(a,a_{0})+\Int{\d 1}
    \left[(h_{A}^{0})^{\text{nn}}(a,1) +
        (h_{A}^{0})^{\text{cn}}(a,1)\right]
    \rho_{A}^{0}(1)F_{A}^{0}(1,a_{0})\notag\\
    &= \left(F_A^0+(h_{A}^{0})^{\text{nc}}+(h_{A}^{0})^{\text{cc}} \right)
    (a,a_0)
\label{4.17}
\end{align}
In $G^{0}_{A}$ all functions are evaluated for the single semi-infinite
plasma $A$ at $\k=0$ and $\k$ has been omitted from the notation.  The
expression for $G_{B}^{0}$ is built in the same way. By the same
calculation that led to (\ref{3.19}) one finds now from (\ref{2.4a}) that
\begin{align}
    \lim_{d\to\infty}d^3\,\avg{f}(d)=\frac{\zeta(3)}{8\pi\beta}
    \left(\frac{\Int{\d a} e_{\alpha_{a}}\,
        \rho_{A}^{0}(a)\,G^{0}_{A}(a,a_{0})}{e_{\alpha_{0}}}\right)
    \left(\frac{\Int{\d b} e_{\beta_{b}}\, \rho_{B}^{0}(b)\,
        G^{0}_{B}(b_{0},b)}{e_{\beta_{0}}}\right)
\label{4.19}
\end{align}
It remains to see that both parenthesis are equal to $-1$ because of the
electroneutrality sum rule in semi-infinite plasmas.  Indeed, using
(\ref{eq:hr}) and (\ref{4.1}), one recognizes from (\ref{4.17}) that
\begin{align}
    G_{A}^{0}(a,a_{0})=h_{A}^{0}(a,a_{0})-\Int{\d 1} \left[
        F_{A}^{0}(a,1)\rho_{A}^{0}(1)+ \delta
        (a,1)\right](h_{A}^{0})^{nn}(1,a_{0}).
\label{4.20}
\end{align}
The contribution to the force of the second term of (\ref{4.20}) involves
\begin{align}
    &\Int{\d a} e_{\alpha_{a}}\,
    \rho_{a}^{0}(a)\left[F_{A}^{0}(a,1)\rho_{A}^{0}(1) +
        \delta(a,1)\right]\nonumber\\
    &=e_{\alpha_{1}}\rho(\alpha_{1}\,x_{1})
    \left(1-\frac{1}{4\pi}\Integral{-\infty}{0}{\d
        x_{a}}(\kappa_{A}^{0})^{2}(x_{a}) \Phi^{0}_{A}(x_{a},x_{1},
    \k=\0)\right)=0
\label{4.21}
\end{align}
which vanishes because of the sum rule (\ref{3.3}) in the case of a
semi-infinite plasma.  The contribution of the first term of (\ref{4.20})
is
\begin{align}
    \Int{\d a} e_{\alpha_{a}}\, \rho^{0}_{A}(a)\, h^{0}_{A}(a,a_{0})=
    \sum_{\alpha_{a}}\!\!\Integral{-\infty}{0}{\d x_{a}}
    e_{\alpha_{a}}\,\rho_{A}^{0}(\gamma_{a}\,x_{a})
    h^{0}_{A}(\alpha_{a}\,x_{a},\alpha_{0}\,0,\k=\0)=-e_{\alpha_{0}}
\label{4.22}
\end{align}
The left-hand side is the total charge of the screening cloud induced in
the semi-infinite plasma $\Lambda_{A}$ by the boundary charge
$e_{\alpha_{0}}$, which equals $-e_{\alpha_{0}}$ because of perfect
screening \cite{martin}. By the same considerations the second parenthesis
in (\ref{4.19}) also equals $-1$, hence the final result (\ref{4.12}).

\section{Plasma in front of a macroscopic dielectric medium}

In this section we investigate the situation where plasma $B$ is replaced
by a semi-infinite macroscopic medium of homogeneous dielectric constant
$\epsilon$. The electrostatic potential $V(\r,\r')$ at $\r$ created by a
unit charge at $\r'\in\LA$ is the Green function of the Poisson equation
with the conditions that the normal component of ${\bf D}(x)=\epsilon
(x){\bf E}(x)$, $(\epsilon(x)=\epsilon,\,x\geq d,\;\epsilon(x)=1,\,x<d)$
and the longitudinal component of ${\bf E}(x)$ are continuous at the
interface \cite{electrodynamics}
\begin{align}
    V(\r,\r') = \begin{cases}
        1/|\r-\r'| + \Delta/|\r-{\r'}^{\ast}|, & x<d,\quad
        \Delta=(1-\epsilon)/(1+\epsilon)\\
        \widetilde{\Delta}/|\r-\r'|, & x>d,\quad
        \widetilde{\Delta}=2/(1+\epsilon)
    \end{cases}
\label{eq:g}
\end{align}
where ${\r}^\ast=(2d-x,\y)$ is the point symmetric to $\r$ with respect to
the dielectric surface. The case of an ideal grounded conducting plate
($\epsilon=\infty$) is formally recovered when $\Delta=-1$,
$\widetilde{\Delta}=0$. In linear electrostatics, the total energy
associated to a distribution of charges $\hat{c}(\r)$ external to the
dielectric is
\begin{align}
    \frac{1}{2}\Int{\d\r} {\bf E}(\r)\cdot {\bf
        D}(\r)=\frac{1}{2}\Int{\d\r}\!\!\!\Int{\d\r'} \hat{c}(\r)V(\r,\r')
    \hat{c}(\r').
\end{align}
For a configuration of charges $\{e_{\alpha_{i}},\r_{i}\}$ in
$\Lambda_{A}$, $\hat{c}(\r)=\sum_{i}e_{\alpha_{i}}\delta (\r-\r_{i}) $, one
finds with (\ref{eq:g}) that the total energy can be written as
\begin{align}
    &U = U_A + U_{AB},\label{6.2} \\
    &U_{A} = \sum_{\{i,j\}} \frac{e_{\alpha_i}
        e_{\alpha_j}}{|\r_i-\r_j|}+\vSR(\alpha_i,\alpha_j,|\r_i-\r_j|),\quad
    U_{AB}= \frac{\Delta}{2}\sum_{i=1}^N \sum_{j=1}^N \frac{e_{\alpha_i}
        e_{\alpha_j}}{|\r_i-\r_j^\ast|}.\notag
\end{align}
where we have omitted the (infinite) self-energies of the particles and
added a short-range repulsive potential for thermodynamic stability. As in
(\ref{1.3}), $U_{AB}$ refers to the additional energy due to the presence
of the dielectric at distance $d$.  The total force exerted by the
dielectric on the particles of $A$ is obtained by differentiating $U$ with
respect to $d$
\begin{align}
    F_{B \to \Lambda_{A} } = \frac{\p}{\p d}U_{AB} = \Delta
    \sum_{i=1}^N\sum_{j=1}^N e_{\alpha_i}
    e_{\alpha_j} \frac{x_i-x_j^\ast}{|\r_i-\r_j^\ast|^{3/2}}.
\end{align}
It corresponds to the sum of all pairwise forces between charges in the
plasma $A$ and their image-charges $\Delta e_{\alpha_j}$. Proceeding as in
the derivation leading from (\ref{1.4}) to (\ref{1.11}), the average force
along the $x$ direction per unit surface is given by
\begin{align}
    \avg{f} &= \lim_{L\to\infty} \frac{\avg{F_{B\to \LA}}_{L}}{L^2}=\Delta
    \Integral{-\infty}{0}{\d x} \Integral{-\infty}{0}{\d x'} \Int{\d \y}
    \frac{x-{x'}^{\ast}}{|\r-{\r'}^{\ast}|^{3/2}} S(x,x',\y) \notag \\
    &= \frac{-\Delta}{d^2} \Integral{-\infty}{0}{\d
        x}\Integral{-\infty}{0}{\d x'} \Int{\frac{\d \q}{(2\pi)^2}} 2\pi
    \e^{-\frac{q}{d}|x+x'|}\e^{-2q}S(x,x',\frac{\q}{d}).
\label{eq:fdiel}
\end{align}
To obtain the second line we have introduced the Fourier transform
(\ref{1.9}), used ${x'}^\ast=2d-x'$, and set $\k=\q/d$. $S(x,x',\y)$ is the
truncated charge-charge correlation function of the plasma $A$ defined in
terms of the statistical weight $\exp(-\beta U)$ associated to the energy
(\ref{6.2}). The asymptotic analysis of (\ref{eq:fdiel}) differs from that
of the previous section on two points~: here the function S provides a
contribution from coincident points to the integrals and we expect
$S(x,x',\tfrac{\q}{d},d)$ to tend towards a non-zero limit when
$d\to\infty$, namely
\begin{align}
    \lim_{d\to\infty}S(x,x',\tfrac{\q}{d})=S^0(x,x',\mathbf{0})
\label{6.3}
\end{align}
with $S^0(x,x',\mathbf{0})$ the charge correlation of the semi-infinite
plasma in absence of the dielectric. Hence, the leading behaviour of the
force as $d\to\infty$
\begin{align}
    \avg{f}\sim \avg{f}^{\text{mon.}} + \avg{f}^{\text{dip.}}
\label{6.4}
\end{align}
comes from the first two terms resulting from the expansion of
$\exp\{-q|x+x'|/d\}\sim 1-q|x+x'|/d$ in (\ref{eq:fdiel}). At leading order
in $ \avg{f}^{\text{dip.}}$ one can replace $S(x,x',\tfrac{\q}{d})$ by
$S^0(x,x',\mathbf{0})$ so that
\begin{align}
    \lim_{\d\to\infty}d^3\avg{f}^{\text{dip.}} &= -\Delta
    \Integral{0}{\infty}{\d q} q^2\e^{-2q}
    \Integral{-\infty}{0}{\d x}\Integral{-\infty}{0}{\d x'}
    (x+x')S^0(x,x',\mathbf{0}) =
    \frac{-\Delta}{16\pi\beta}.
\label{6.5}
\end{align}
The first term of the $x,x'$ integrals vanishes because of perfect
screening whereas the second one equals $1/(4\pi\beta)$ as a consequence of
dipole sum rule in a semi-infinite plasma (\cite{martin}, formula (3.9),
sect. C)
\footnote{Here the sign is opposite to that in \cite{martin} because the
    plasma is located in the $x<0$ half space.}.  Since
$\lim_{d\to\infty}d^2 \avg{f}^{\text{mon.}} = 0$ because of perfect
screening, one can replace $S(x,x',\tfrac{\q}{d})$ by
$S(x,x',\tfrac{\q}{d})-S^0(x,x',\mathbf{0})$ in the monopole contribution
$\avg{f}^{\text{mon.}}$. It is convenient to further add and subtract
$S^0(x,x',\tfrac{\q}{d})$ and to note that~:
\begin{align}
    \lim_{d\to\infty}d^3
    &\left\{\frac{-\Delta}{d^2}\Integral{-\infty}{0}{\d x}
    \Integral{-\infty}{0}{\d x'} \Integral{0}{\infty}{\d q} q\e^{-2q}
    \left[S^0(x,x',\tfrac{\q}{d}) - S^0(x,x',\mathbf{0})\right]\right\}\\
    &= -\Delta \Integral{0}{\infty}{\d q} q^{2}\e^{-2q}
    \left. \tfrac{\d}{\d k} \right\vert_{k=0}
    \left(\Integral{-\infty}{0}{\d x}
    \Integral{-\infty}{0}{\d x'} S^0(x,x',k)\right) =
    \frac{-\Delta}{16\pi\beta}.
\label{6.6}
\end{align}
This follows from the small $k=|\k|$ expansion of the $x,x'$ integral that
has a linear term $\frac{k}{4\pi\beta}$ (\cite{martin}, formula (3.24),
sect. C).  Collecting (\ref{6.5}) and (\ref{6.6}) in (\ref{6.4}) we see
that the large-$d$ behaviour of $\avg{f}$ is determined by
\begin{align}
    \avg{f} = \frac{-\Delta}{8\pi\beta d^3} - \frac{\Delta}{d^2}
    \Integral{-\infty}{0}{\d x}
    \Integral{-\infty}{0}{\d x'} \Integral{0}{\infty}{\d q} q\e^{-2q}
    \left[S(x,x',\tfrac{\q}{d}) - S^0(x,x',\tfrac{\q}{d}) \right] +
    \text{o}(d^{-3}).
\label{eq:force-diel}
\end{align}
One can now proceed as in section \ref{sec:mayer} with Mayer bonds defined
in terms of the Green function (\ref{eq:g}).  The Debye-H\"uckel equation
(\ref{eq:mfpotential-diffeq}) is supplemented with the boundary condition
$\lim_{x\to d_{-}}\p \Phi(x,x',\k)/ \p x =\epsilon\,\lim_{x\to d_{+}}\p
\Phi(x,x',\k)/ \p x$, and $\kappa(x)$ is as before for $x\leq 0$, and
$\kappa(x)=0,\;x>0$. The solution for piecewise-flat densities is
(\cite{aqua-cornu-profiles1}, formulae (3.2)-(3.5))~\footnote{Here
    the minimal distance $d$ between a charge and the dielectric wall plays
    the role of the hard-core diameter in
    \cite{aqua-cornu-profiles1}. Notice that we defined $\Delta$ with the
    opposite sign and that our plasma fills the region $x<0$.}
\begin{align}
    \varphi(x,x',\k) = 2\pi \frac{\e^{-k_A|x-x'|}}{k_A} +
    2\pi\frac{\e^{-k_A|x+x'|}}{k_A} \frac{(k_A-k)\e^{kd}+\Delta(k_A+k)
        \e^{-kd}}{(k_A+k) \e^{kd} + \Delta(k_A-k)\e^{-kd}}.
\label{eq:varphi-diel}
\end{align}
It is convenient to single out the potential $\varphi^0(x,x',\k)$
(\ref{3.6}) for the semi-infinite plasma in the absence of the dielectric
and to split
\begin{align}
    &\varphi(x,x',\k) = \varphi^0(x,x',\k) +
    \varphi_{AB}(x,x',\k),\label{eq:splitting}
\end{align}
where, from (\ref{3.6}) and (\ref{eq:varphi-diel}),
\begin{align}
    &\varphi_{AB}(x,x',\k) = \frac{8\pi k \e^{-k_A|x|} \e^{-k_A|x'|} \Delta
        \e^{-kd}}{(k_A+k)\left[(k_A+k)\e^{kd}+(k_A-k)\Delta
            \e^{-kd}\right]}
\label{6.7}
\end{align}
One observes that $\varphi_{AB}(x,x',\k)$ has the factorization property
analogous to (\ref{3.7})
\begin{align}
    &\lim_{d\to\infty} d \varphi_{AB}(x,x',\tfrac{\q}{d}) = z(q)
    \varphi^0(x,0,\mathbf{0})
    \varphi^0(0,x',\mathbf{0}),\label{eq:factorization-diel}\\
    &z(q) = \frac{q}{2\pi}\,\frac{\Delta \e^{-2 q}}{1+\Delta \e^{-2q}}.
\label{eq:z-diel}
\end{align}
The potential $\Phi$ corresponding to the exact non-uniform profile is
related to $\varphi$ by the integral equation
(\ref{eq:mfpotential-integraleq2}). With a reasoning similar to that
leading to (\ref{3.16}) and (\ref{eq:mfpotential-almost-isol-parts}), it
can also be split in two parts
\begin{align}
    \Phi(x,x',\k) \equiv \widetilde{\Phi}(x,x',\k) + \Phi_{AB}(x,x',\k)
\label{6.8}
\end{align}
Here $\widetilde{\Phi}$ verifies
eq. (\ref{eq:mfpotential-almost-isol-parts}) with $\varphi^{0}$ in place of
$\varphi_{AA}$ and tends to the potential $\Phi^{0}$ of the semi-infinite
plasma in vacuum.  $\Phi_{AB}$ solves eq. (\ref{3.16}) with $u$ in place of
$u_{A}$ and $u_{B}$, $\widetilde{\Phi}$ in place of
$\widetilde{\Phi}_{AA}$, and $\Phi$ in place of $\Phi_{BB}$.  Then using
(\ref{eq:factorization-diel}) and the fact that both $\Phi$ and
$\widetilde{\Phi}$ tend to $\Phi^{0}$ leads to the asymptotic factorisation
of $\Phi_{AB}$
\begin{align}
    \lim_{d\to\infty} d \Phi_{AB}(x,x',\tfrac{\q}{d}) = z(q)
    \Phi^0(x,0,\mathbf{0}) \Phi^0(0,x',\mathbf{0}).
\label{6.9}
\end{align}
We are now ready to evaluate the force (\ref{eq:force-diel}) as
$d\to\infty$ in the Debye-H\"uckel approximation. From (\ref{2.4}),
(\ref{eq:kappa}) and (\ref{3.1}) one has
\begin{align}
    S^{{\rm DH}}(x,x',\tfrac{\q}{d})=
    -\frac{1}{\beta}\frac{\kappa^{2}(x)}{4\pi} \frac{\kappa^{2}(x')}{4\pi}
    \Phi(x, x',\tfrac{\q}{d}) + \delta (x-x')
    \frac{\kappa^{2}(x)}{4\pi\beta}
\label{6.10}
\end{align}
and the analogous relation for $S^{0\,{\rm DH}}(x,x',\tfrac{\q}{d})$; some
care has to be exercised here since coincident points do contribute when
both $x,x'$ are in the same integration range. We subtract and add
$\widetilde{\Phi}(x,x',\tfrac{\q}{d})$ to $\Phi(x,x',\tfrac{\q}{d})$ in
(\ref{6.10}). This gives two contributions to the force
(\ref{eq:force-diel}). The first one is
\begin{align}
    &\frac{-\Delta}{8\pi\beta d^3} + \frac{\Delta}{\beta d^2}
    \Integral{-\infty}{0}{\d x} \Integral{-\infty}{0}{\d x'}
    \Integral{0}{\infty}{\d q} q\,\e^{-2q}\,
    \frac{\kappa^{2}(x)}{4\pi}\frac{\kappa^{2}(x')}{4\pi}
    \left[\Phi(x,x',\tfrac{ \q}{d})
        -\widetilde{\Phi}(x,x',\tfrac{\q}{d})\right] \nonumber\\
    &=\frac{-\Delta}{8\pi\beta d^3} + \frac{\Delta}{\beta d^2}
    \Integral{-\infty}{0}{\d x} \Integral{-\infty}{0}{\d x'}
    \Integral{0}{\infty}{\d q} q\,\e^{-2q}\,
    \frac{\kappa^{2}(x)}{4\pi}\frac{\kappa^{2}(x')}{4\pi}
    \Phi_{AB}(x,x',\tfrac{\q}{d})
\label{6.11}
\end{align}
The second one involves the quantity
\begin{align}
    &\frac{1}{4\pi\beta}\left[\kappa^2(x)-(\kappa^0)^2(x)\right] \notag\\
    &\phantom{=}+\frac{-1}{\beta} \Integral{-\infty}{0}{\d x'}
    \biggl[\frac{\kappa^2(x)}{4\pi}
        \frac{\kappa^2(x')}{4\pi}\widetilde{\Phi}(x,x',\tfrac{\q}{d}) -
        \frac{(\kappa^0)^2(x)}{4\pi}\frac{(\kappa^0)^2(x')}{4\pi}
        \Phi^0(x,x',\tfrac{\q}{d})\biggr]\nonumber\\
    &=\frac{q^2}{4\pi\beta d^{2}} \Int{\d x'}
    \left[\widetilde{\Phi}(x,x',\tfrac{\q}{d}) -
        \Phi^0(x,x',\tfrac{\q}{d})\right]={\cal
        O}\left(\frac{1}{d^{2}}\right)
\label{eq:Stilde-S0-diel}
\end{align}
This equality follows from the relation (\ref{3.3a}) for $\widetilde{\Phi}$
(relative to $\kappa$) and for $\Phi^{0}$ (relative to $\kappa^{0}$) since
both potentials satisfy the basic differential equation
(\ref{eq:mfpotential-diffeq}). Thus the contribution of
(\ref{eq:Stilde-S0-diel}) to the force is ${\cal
    O}\left(\frac{1}{d^{4}}\right)$.  With the factorisation (\ref{6.9})
and using the sum rule (\ref{3.3}) for $\Phi^{0}$, one finds from
(\ref{6.11}) and (\ref{eq:z-diel}) the final result
\begin{align}
    \lim_{d\to\infty} d^3 \avg{f} = \frac{-\Delta}{8\pi\beta} +
    \frac{\Delta}{\beta} \Integral{0}{\infty}{\d q} q\,\e^{-2q}\,z(q)
    =\frac{1}{8\pi\beta} \sum_{n=1}^\infty \frac{(-\Delta)^n}{n^3}
\label{eq:f-diel}
\end{align}
It can be shown along the lines presented in section 5 that the non
mean-field part of the charge correlation function does not contribute to
this asymptotic result.  According to (\ref{6.8}) one splits the bonds
$F=\widetilde{F}+ F_{AB}$ and $F^{R}=\widetilde{F}^{R}+F^{R}_{AB}$ with
$F^R_{AB}\equiv \exp({F_{AB}}-\beta v_{SR}) -1-F_{AB}$. The bonds
$\widetilde{F}$ and $\widetilde{F}^{R}$ tend to the bonds $F^{0}$ and
$F^{0\,R}$ pertaining to the semi-infinite plasma without dielectric,
whereas $F_{AB}$ vanishes in the limit. At large separation, $F^R_{AB}\sim
(F_{AB})^{2}$ vanishes more rapidly than $F_{AB}$ (see appendix A).  As in
the analysis leading to (\ref{4.16}), the leading behaviour of the Ursell
function comes from graphs having bonds $\widetilde{F}$,
$\widetilde{F}^{R}$, a single $F_{AB}$ one, and it takes the factorized
form (\ref{4.16}). The only difference is that both functions $G^{0}$ refer
to the same plasma $A$.  Then one establishes the validity of
(\ref{eq:f-diel}) as in (\ref{4.19})-(\ref{4.22}).

This result coincides with that of Lifshitz. Indeed, a straightforward
generalisation of its asymptotic force as $\frac{d}{\beta \hbar c}\gg 1$
(formula (5.5) in \cite{lifshitz1}) to the case of two
different homogeneous dielectric media of constants $\epsilon_1,
\epsilon_2$ yields~\footnote{
One generalises formulae (5.2), (5.3)
    and finally (5.5) of \cite{lifshitz1} starting from (2.4) by keeping
$\epsilon_1$ and $\epsilon_2$ different.}
\begin{align}
    f \sim \frac{1}{8\pi\beta d^3} \sum_{n=1}^\infty \frac{(\Delta_1
        \Delta_2)^n}{n^3},\qquad
    \Delta_i=\frac{1-\epsilon_i}{1+\epsilon_i},\ i=1,2.
\label{eq:lifshitz}
\end{align}
This reduces to (\ref{eq:f-diel}) once one of the slabs is a conductor, i.e
$\Delta_1=-1$.  We have therefore provided a derivation of this formula
when the conductor is described as a statistical system of fluctuating
charges in thermal equilibrium. It is interesting to note that thermal
fluctuations in one of the slabs suffice to generate the correct asymptotic
value of the force.

\subsubsection*{Acknowledgements}
We thank F. Cornu and B. Jancovici for useful discussions.

\appendix
\section{Slab of finite thickness}

The analysis of sections 2-5 applies to infinitely thick slabs. It is
interesting to check that the asymptotic behaviour of the force does not
depend on the slab thickness. We consider now that the slab $\Lambda_{A}$
has finite thickness $a<\infty$, while, for simplicity, we keep the slab
$\Lambda_{B}$ infinitely thick.  The setting of sections 2 and 3 remains
the same with the $x-$integration on $\Lambda_{A}$ limited to the interval
$-a\leq x\leq 0$. We then follow the same route as in section 4 and 5 by
first considering the equation (\ref{3.5}) for the piecewise-flat profile
\begin{align}
    &\bar{\kappa}(x)=0,\; x<-a,\quad
    \bar{\kappa}(x)=\frac{1}{a}\Integral{-a}{0}{\d x}
    \kappa(x)\equiv\kappa_{A}, \;-a<x<0\nonumber\\
    &\bar{\kappa}(x)=0,\;0<x<d,\quad\bar{\kappa}(x)=\kappa_{B},\;x>d
\label{C.1}
\end{align}
It is convenient to choose $\bar{\kappa}(x)$ equal to the average of
$\kappa(x)$ in $\Lambda_{A}$, since we expect the latter to be close to its
mean value at weak-coupling.

Solving equation (\ref{3.5}) with $x'$ fixed, continuity conditions of
$x\mapsto\varphi(x,x',\k)$ and of $x\mapsto\p_x\varphi(x,x',\k)$ at
$x=-a,x=0,x=d$ as well with $\lim_{x\to\pm\infty}\varphi(x,x',\k) = 0$
yields
\begin{align}
    &\varphi(x,x',\k)=\begin{cases}
    	\varphi_{AA}(x,x',\k), \quad &-a<x,x'<0\\
        \varphi_{AB}(x,x'-d,\k), \quad &-a<x<0<d<x'\\
        \varphi_{BB}(x-d,x'-d,\k), \quad &d<x,x'
        \end{cases}\label{eq:phi-cases}
\end{align}
where
\begin{align}
    &\varphi_{AA}(x,x',\k) = \tfrac{2\pi}{k_A}\tfrac{(k_A+k)\e^{k_A a}
        \left( \e^{-k_A|x'-x|}\sigma_1 + \e^{-k_A|x'+x|}\sigma_2\right) +
        (k_A-k)\e^{-k_A a} \left(\e^{k_A|x'-x|}\sigma_2 +
        \e^{k_A|x'+x|}\sigma_1\right)}{(k_A+k)\e^{k_A
            a}\sigma_1-(k_A-k)\e^{-k_A a}\sigma_2} \label{eq:phiAA-a}\\
    &\varphi_{AB}(x,x',\k) = \frac{8\pi k\left[(k_A+k)\e^{k_A a}\e^{k_A x}
            + (k_A-k)\e^{-k_A a}\e^{-k_A x}\right] \e^{-k_B
            x'}}{(k_A+k)\e^{k_A a} \sigma_1 - (k_A-k)\e^{-k_A
            a}\sigma_2} \label{C.2}\\
    &\varphi_{BB}(x,x',\k) =
    \frac{2\pi}{k_B}\left(\e^{-k_B|x'-x|}+\e^{-k_B|x'+x|}
    \tfrac{(k_A+k)\e^{k_A a}\sigma_3-(k_A-k)\e^{-k_A
            a}\sigma_4}{(k_A+k)\e^{k_A a}\sigma_1-(k_A-k)\e^{-k_A
            a}\sigma_2}\right)\label{C.3}
\end{align}
and
\begin{alignat}{2}
    &\sigma_1={\scriptstyle (k_A+k)(k_B+k)\e^{kd}-(k_A-k)(k_B-k)\e^{-kd}}
    &\quad
    &\sigma_2={\scriptstyle
        (k_A-k)(k_B+k)\e^{kd}-(k_A+k)(k_B-k)\e^{-kd}}\label{C.4}\\
    &\sigma_3={\scriptstyle (k_A+k)(k_B-k)\e^{kd}-(k_A-k)(k_B+k)\e^{-kd}}
    &\quad
    &\sigma_4={\scriptstyle
        (k_A-k)(k_B-k)\e^{kd}-(k_A+k)(k_B+k)\e^{-kd}}\nonumber\\
    &k_A = \sqrt{k^2+\KA^2} &\quad &k_B = \sqrt{k^2+\KB^2}\nonumber.
\end{alignat}
One deduces also from the differential equations that for any $a>0$ both
$\varphi$ and $\Phi$ verify the charge sum rule (\ref{3.3}).  As
$a\to\infty$, formulae reduce to those obtained in section 4 for two
semi-infinite plasmas.

The main observation to be made on this explicit result is that it obeys
exactly the same factorisation property in terms of the scaled variable
$\q=\k d$ as (\ref{3.7}) with the the same factors (here
$\varphi_{A}^{0}(x,0,\0)$ corresponds to the single plasma $A$ with finite
thickness). The rest of the analysis is the same as in section 4, with the
difference that the above solution verifies $a-$dependent bounds in place
of (\ref{eq:mfflatpotential-bounds})-(\ref{3.9}), namely (appendix B)
\begin{align}
    \left|\varphi(x,x',\k)\right| \leq \varphi^>(x,x') \leq
    \frac{4\pi}{\kappa} \coth \kappa a,
    \quad \kappa d \geq 1,
\label{C.5}
\end{align}
where $\varphi^>(x,x')$ is defined piecewise from $\varphi_{AA}^>$,
$\varphi_{AB}^>$, etc. as in (\ref{eq:phi-cases}) with
\begin{align}
    &\varphi_{AA}^>(x,x') = \frac{2\pi}{\kappa}\, \frac{\cosh
        \kappa(a-|x'-x|) + \cosh \kappa(a-|x'+x|)}{\sinh \kappa
        a}\label{C.6}\\
    &\varphi_{AB}^>(x,x') = \frac{4\pi}{\kappa^2 d}\, \frac{\cosh
        \kappa(a-|x|)}{\sinh \kappa a}\ \e^{-\kappa x'}\label{C.7}\\
    &\varphi_{BB}^>(x,x') =
    \frac{2\pi}{\kappa}\left(\e^{-\kappa|x'-x|}+\e^{-\kappa|x'+x|}\right)
    , \quad\kappa := \min\{\KA,\KB\}\label{C.8}
\end{align}
The potential $\Phi$ with structured profiles is related to $\varphi$ by
the integral equation (\ref{eq:mfpotential-integraleq2}) and the estimate
(\ref{eq:mfpotential-bounds}) of the lemma becomes
\begin{align}
    \left|\Phi(x,x',\k)\right| \leq \frac{1}{1-r(a)}\ \varphi^>(x,x'),
    \quad r(a) := r\, \coth \kappa a
\label{C.9}
\end{align}
where $r$ is defined by (\ref{3.10}). To have $r(a)<1$ one needs $r$
sufficiently small (weak-coupling, see appendix B) and $\kappa a$ not too
small, i.e. the slab width is larger than the typical screening length in
the plasma
\footnote{ Notice that the above bounds cannot be uniform in $a$~:
    $\varphi_A^0(x,x',\mathbf{0})$ diverges as $a\to 0$ so that its
    integral over $[-a,0]$ leads to the constant value $\frac{4\pi}{\KA}$
    requested by the charge sum rule.}.  Then the steps leading to
(\ref{3.19}) are the same as in section 4 and the considerations of section
5 apply as well.  The reason for the asymptotic force being independent of
the slab thickness is clearly displayed in expressions (\ref{3.19}) and
(\ref{4.19}): it only depends on the screening cloud associated to charges
located at the inner boundaries of the slabs, and thus is insensitive to
charge fluctuations elsewhere in the slabs.

\section{Bounds for the Debye-H\"uckel potential}
In this appendix we present some details of the calculations leading to
bounds used throughout the paper for the Debye-H\"uckel potentials
$\varphi$ and $\Phi$ and discuss the validity of
(\ref{eq:mfpotential-bounds}) in the weak-coupling regime.

At first, we show the bound (\ref{C.5}), which is a generalisation of
(\ref{eq:mfflatpotential-bounds}) to the case where plasma $A$ is of finite
thickness $a$. Result (\ref{eq:mfflatpotential-bounds}) is recovered by
taking $a\to \infty$.  From (\ref{C.4}), one has $\sigma_1 \geq \sigma_2$;
$\sigma_1 \geq \kappa_A \kappa_B(\e^{kd}-\e^{-kd})$, and
$\sigma_2+\sigma_4\leq\sigma_1+\sigma_3$;
$\sigma_3-\sigma_1\leq\sigma_4-\sigma_2<0$. This implies
\begin{align}
    -1 \leq \tfrac{(k_A+k)\e^{k_A a} \sigma_3 - (k_A-k)\e^{-k_A a}
        \sigma_4}{(k_A+k) \e^{k_A a} \sigma_1 - (k_A-k) \e^{-k_A a}
        \sigma_2} \leq 1,\notag
\end{align}
which yields the bound (\ref{C.8}) for $\varphi_{BB}$~:
\begin{align}
    \left|\varphi_{BB}\right| \leq
    \tfrac{2\pi}{k_B}\left(\e^{-k_B|x'-x|}+\e^{-k_B|x'+x|}\right) \leq
    \varphi_{BB}^>(x,x') .  \notag
\end{align}
To obtain the bound (\ref{C.7}) for $\varphi_{AB}$, we then note that
\begin{align}
    &\sigma_1-\tfrac{k_A-k}{k_A+k}\e^{-2 k_A a} \sigma_2 \geq
    \left(1-\e^{-2 k_A a}\right)\sigma_1.
    \label{eq:sigma-bound}\\
    & \text{Thus,}\;\;\frac{k}{\sigma_1-\frac{k_A-k}{k_A+k} \e^{-2 k_A a}
        \sigma_2} \leq \frac{1}{1-\e^{-2k_A a}} \,\frac{1}{\kappa_A
        \kappa_B d}\, \frac{k d}{\e^{kd}-\e^{-kd}} \leq
    \frac{1}{1-\e^{-2k_A a}} \,\frac{1}{\kappa^2 d}\, \frac{1}{2},\notag
\end{align}
so that
\begin{align}
    0 \leq \varphi_{AB}(x,x',\k) \leq \tfrac{4\pi}{\kappa^2 d}
    \frac{\e^{k_A(a-|x|)}+\e^{-k_A(a-|x|)}}{\e^{k_A a}-\e^{-k_A a}}
    \e^{-k_B x'} \leq \varphi_{AB}^>(x,x').\notag
\end{align}
Last inequality uses
\begin{align}
    \frac{\e^{kx}+\e^{-kx}}{\e^{kX}-\e^{-kX}} \leq \frac{\e^{\kappa x} +
        \e^{-\kappa x}}{\e^{\kappa X}-\e^{-\kappa X}}, \qquad
    \text{$0<x\leq X$, $\ k\geq \kappa >0$}.
    \label{eq:ch-sh-bound}
\end{align}
Finally, (\ref{eq:sigma-bound}), the fact that $|\sigma_2/\sigma_1| \leq 1$
and (\ref{eq:ch-sh-bound}) show the bound (\ref{C.6})
\begin{align}
    \left|\varphi_{AA}(x,x',\k)\right| &\leq \tfrac{2\pi}{\kappa_A}
    \frac{\sigma_1 \e^{-k_A|x'-x|} + |\sigma_2| \e^{-k_A|x'+x|} + \e^{-2
            k_A a}\left(|\sigma_2|\e^{k_A |x'-x|} + \sigma_1
        \e^{k_A|x'+x|}\right)}{\left(1-\e^{-2k_A a}\right) \sigma_1}
    \notag\\
    &\leq \varphi_{AA}^>(x,x').\notag
\end{align}

\subsubsection*{Proof of the lemma}
To proof the bound (\ref{eq:mfpotential-bounds}) of $\Phi(x,x',\k)$, we
proceed as follows. By (\ref{eq:mfpotential-integraleq2}) we can develop
$\Phi(x,x',\k)$ as a perturbation series w.r.t. $\varphi(x,x',\k)$, whose
$n^{\text{th}}$ term reads
\begin{align}
    \left(\frac{-1}{4\pi}\right)^n \Int{\d s_1\cdots\d s_n} u(s_1)\cdots
    u(s_n) \varphi(x,s_1,\k) \varphi(s_1,s_2,\k) \cdots \varphi(s_n,x',\k).
\end{align}
This term is bounded by $r^n \varphi^>(x,x')$, where $r$ is given by
(\ref{3.10}). Indeed, according to (\ref{eq:mfflatpotential-bounds}),
$\varphi(x,x',\k)$ is bounded by $\varphi^>(x,x')$, which itself satisfies
\begin{align}
    \varphi^>(x,s) \varphi^>(s,x') \leq \frac{4\pi}{\kappa}
    \varphi^>(x,x'), \quad \forall s, x, x' \label{eq:varphi-contraction}.
\end{align}

Consequently, if $r<1$, the series is absolutely convergent and the lemma
holds.  Inequality (\ref{eq:varphi-contraction}) is proven using
(\ref{3.8}), (\ref{3.9}) and verifying it for each case. As an example,
\begin{align}
    \varphi_{AA}^>(x,s)\varphi_{AB}^>(s,x') &=
    \frac{2\pi}{\kappa}\,\frac{4\pi}{\kappa^2 d} \e^{-\kappa |x'|} \left(
    \e^{-\kappa \left(|x-s|+|s|\right)} + \e^{-\kappa
        \left(|x+s|+|s|\right) } \right)\\
    &\leq \frac{4\pi}{\kappa}\,\varphi_{AB}^>(x,x'),
\end{align}
because $|x\pm s| + |s| \geq |x|$. Some of the majorations leading to
(\ref{eq:varphi-contraction}) assume $\kappa d \geq 1$.

If plasma $A$ is finitely-thick, (\ref{eq:varphi-contraction}) generalizes
to
\begin{align}
    \varphi^>(x,s)\varphi^>(s,x') \leq \frac{4\pi}{\kappa}
    \coth(\kappa a) \ \varphi^>(x,x'),\quad \forall x,x',s
    \label{eq:varphi-contraction2}
\end{align}
and we obtain the bound (\ref{C.9}) for $\Phi$.

\subsubsection*{Profiles in weakly-coupled plasmas}
The parameter $r$ occurring in the bound (\ref{eq:mfpotential-bounds}) can
be chosen small enough in the weak-coupling regime, defined by $\Gamma =
\tfrac{1}{2}\beta e^2 \kappa \ll 1$ ($e$ is a typical charge of the
system). Indeed, to estimate $r$, the deviations of the density profiles to
their bulk values need to be known. In the simplest case of a semi-infinite
charge-symmetric plasma \footnote{A charge-symmetric plasma has two species
    with opposite charges and same bulk densities.}  in the weak-coupling
regime, Jancovici \cite{jancovici-along-the-wall} finds
\begin{align}
    \frac{\rho_A(\gamma\, x)-\rho_{A\,\gamma}}{\rho_{A\,\gamma}} \simeq
    \frac{\beta e_\gamma^2}{2}\, \kappa_A \chi(\kappa_A x),\quad \Gamma\ll
    1\notag,
\end{align}
where $\chi$ is integrable. Integrating over $x<0$ and forming $r$
according to (\ref{3.10}) shows that $r$ is proportional to $\Gamma$. This
also holds if the semi-infinite plasma is not charge-symmetric (see
\cite{aqua-cornu-profiles1}, sect. 5). For our two-plasmas system, $r$ will
be less than $1$ provided $\Gamma$ is small and $d$ is large.

\section{Decay of Mayer graphs at large slab separation}

We consider prototype graphs constituted of bonds (\ref{4.13}) labelled by
the indices $AA, AB, BA, BB$ according to the respective location of the
variables $x,x'$ in slab $A$ or in slab $B$. In view of (\ref{2.4a}) and
after the changes $\k d=\q,\; x'\to x'-d$, the contribution to the force of
a graph $\Pi_{AB}$ with first root point in $\Lambda_{A}$ and second root
point in $\Lambda_{B}$ is
\begin{align}
    \avg{f}^{\Pi_{AB}} = -\frac{1}{d^2} \Integral{-\infty}{0}{\d x}\!\!
    \Integral{0}{\infty}{\d x'} \!\!\!\!\Integral{0}{\infty}{\d q} q
    \e^{-\frac{q}{d}|x-x'|}\e^{-q} \sum_{\gamma,\gamma'} e_\gamma
    e_{\gamma'} \rho_A(\gamma\,x)\rho_B(\gamma'\,x')
    \pi_{AB}(\gamma\,x,\gamma'\,x',\tfrac{\q}{d}).
\label{B.0}
\end{align}
A graph $\Pi_{AB}$ having $L$ bonds of $\mathcal{F}_{AB}$ or
$\mathcal{F}_{BA}$ type, written in Fourier space with respect to the
$\y$-variables, is of the general form
\begin{align}
    &\Pi_{AB}(\gamma\,x,\gamma'\,x',\k) = \frac{1}{S_{\pi_{AB}}} \Int{\d 1}
    \rho(1) \cdots \Int{\d m} \rho(m) \Int{\prod_{j=1}^L \frac{\d
            \pmb{\ell}_j}{(2\pi)^2} \mathcal{F}_{[AB]}(\pmb{\ell}_j)}
    \notag\\
    &\times\Int{\prod_a \frac{\d \k_a}{(2\pi)^2} \mathcal{F}_{AA}(\k_a)}
    \Int{\prod_b \frac{\d \k_b}{(2\pi)^2}\mathcal{F}_{BB}(\k_b)}
    \prod_{n=0}^m (2\pi)^2
    \delta[\k,\{\pmb{\ell}_{j}\}_{n},\{\k_a\}_{n},\{\k_b\}_{n}].
\label{B.1}
\end{align}
Here $m$ is the number of internal points, $i=(\gamma_i,x_i)$ and $\Int{\d
    i}$ stands for summation over particle species and integration over
$x_i$.  $\mathcal{F}(\k)$ stands either for $F(\k)$ or $F^{\text{R}}(\k)$
and we have omitted the species and $x,x'$ dependencies from the notation
\footnote{In (\ref{B.1}), $\mathcal{F}_{[AB]}$ designates either a
    $\mathcal{F}_{AB}$ or a $\mathcal{F}_{BA}$ bond.  Notice that
    $\mathcal{F}_{BA}(i,j,\k)=\mathcal{F}_{AB}(j,i,\k)$. }.  The product of
$m+1$ $\delta$-functions expresses the conservation of wave numbers at the
$m$ internal points plus a relation that fixes the sum of ingoing (or
outgoing) wave numbers to $\k$, as a result of $\y-$translation invariance.
These constitute $m+1$ linear equations between wave numbers from the sets
$\{\pmb{\ell}_{j}\},\{\k_a\},\{\k_b\}$, which imply $C$ independent
relations involving only $\pmb{\ell}$ variables.  Depending on the topology
of the graph, $1\leq C \leq L$.  Consider for instance the graph
constituted by a single chain of bonds
$F_{AB}(\pmb{\ell}_1){F}^{R}_{BB}(\k_{b})
F_{BA}(\pmb{\ell}_2)F^{R}_{AA}(\k_{a})F_{BA}(\pmb{\ell}_3)$ with $L=3\,$
\footnote{The ${F}^{R}$ bonds are needed because of the excluded
    convolution rule between $F$ bonds.}.  The conservation laws
$\k=\pmb{\ell}_{1}=\k_{a}=\pmb{\ell}_{2}=\k_{b}=\pmb{\ell}_{3}$ imply the
independent relations
$\k=\pmb{\ell}_{1},\;\pmb{\ell}_{1}=\pmb{\ell}_{2},\;\pmb{\ell}_{2} =
\pmb{\ell}_{3}$ between the $\pmb{\ell}$ variables, thus $C=3$.  Consider
now the graph constituted of two parallel chains
$F_{AA}^{R}(\k_{a1})F_{AB}(\pmb{\ell}_1)F_{BB}^{R}(\k_{b1})$ and
$F_{AA}^{R}(\k_{a2})F_{AB}(\pmb{\ell}_2)F_{BB}^{R}(\k_{b2})$ with
$L=2$. The conservation laws are $\k=\k_{a1}+\k_{a2},
\;\k_{a1}=\pmb{\ell}_{1}=\k_{b1}, \;\k_{a2}=\pmb{\ell}_{2 }=\k_{b2}$
implying the single relation $\k=\pmb{\ell}_{1}+\pmb{\ell}_{2}$, thus
$C=1$.

Then we perform the integrations in (\ref{B.1}) in the following order. We
first carry $C$ integrals on the $\delta$ functions corresponding to the
above relations between $\pmb{\ell}$ variables: this expresses the
$\pmb{\ell}_{j}$ variables $j=1,\ldots,L$ in the integrand as linear
combinations of the remaining $L-C$ ones, say
$\pmb{\ell}_{C+1},\ldots,\pmb{\ell}_{L}$. We evaluate now $\Pi_{AB}$ at
$\k=\q/d$ and change the variables $\pmb{\ell}_j=\q_{j}/d,
\;j=C+1,\ldots,L$: the Jacobian provides a factor $d^{-2(L-C)}$.  As
$d\to\infty$ the $\k_a,\;\k_{b}$ and $\q$ integrals factorize.  Indeed in
$\mathcal{F}_{AA}$ or $\mathcal{F}_{BB}$ bonds the $\q$ dependences occur
in the form $\q/d\to 0, d\to \infty$ whereas for $\mathcal{F}_{AB}$ or
$\mathcal{F}_{BA}$ bonds we use the asymptotic form (\ref{3.17}). This
yields a factor $d^{-n[AB]}$ if the number of $F_{[AB]}$ bonds is $n[AB]$
and $d^{- 4n^{\text{R}}[AB]}$ if the number of $F^{\text{R}}_{[AB]}$ bonds
is $n^{\text{R}}[AB]$ (for the latter, see at the end of this appendix).
Then the $\k_{a}$ and $\k_{b}$ integrals refer to products of
$\mathcal{F}_{AA}$ and $\mathcal{F}_{BB}$ as in single semi-infinite
plasmas and the $\q$ integrals are carried on product of functions $q/\sinh
q$. For the above examples the $\q$ integrals are $\int d\q (q/\sinh
q)^{3}$ and $\int d\q_{1}\int d\q_{2} (q_{1}/\sinh q_{1})( |\q_{1}-\q_{2}|/
\sinh |\q_{1}-\q_{2}|)$.  As a final result a graph decays as
\begin{align}
    d^{-2(L-C)} d^{-n[AB] } d^{- 4n^{\text{R}}[AB]}, \quad d\to\infty
\label{B.2}
\end{align}
times a factor of order one resulting of the above integrals.  It is clear
that the minimal decay $d^{-1}$ is obtained when there is only one $F_{AB}$
bond.

It remains to examine the decay of a $F^{\text{R}}_{[AB]}$ bond, which
reads in Fourier form, according to (\ref{eq:Fr})
\begin{align}
    F^{\text{R}}_{AB}(\k)=\Int{\d\y} e^{-i\k\cdot\y}\left[\exp(-\beta
        e_{1}e_{2} \Phi_{AB}(\y))-1+ \beta e_{1}e_{2}\Phi_{AB}(\y)\right].
\label{B.3}
\end{align}
The $x$ variables are omitted and $|x_{1}-x_{2}|$ is large enough so that
the short range part of the potential does not contribute. In view of
(\ref{3.17}), $\Phi_{AB}(\y)$ has the asymptotic form
\begin{align}
    \Phi_{AB}(\y)&=\frac{1}{d^{2}} \Int{\frac{\d\q}{(2\pi)^{2}}}
    \exp\left(i\tfrac{\q}{d}\cdot\y\right)\Phi_{AB}
    \left(\tfrac{\q}{d}\right)\nonumber\\
    &\sim\frac{1}{d^{3}}f\left(\tfrac{\y}{d}\right)\quad\quad
         {\text{with}}\quad\quad f(\y)=\Phi^{0}_{A} \Phi^{0}_{B} \Int{
             \frac{\d\q}{(2\pi)^{2}}} e^{i\q\cdot\y}\frac{q}{4\pi\sinh q}.
\label{B.4}
\end{align}
Hence substituting (\ref{B.4}) in (\ref{B.3}) and expanding for large $d$
after the change of variable $\mathbf{u}=\y/d$ gives
\begin{align}
    F^{\text{R}}_{AB}\left(\frac{\q}{d}\right)& \sim d^{2}
    \Int{\d\mathbf{u}} e^{-i\q\cdot\mathbf{u}}
    \left[\exp\left(-\frac{\beta}{d^{3}}f(\mathbf{u})\right) -1
        +\frac{\beta}{d^{3}}f(\mathbf{u})\right]\nonumber\\
    &\sim \frac{1}{d^{4}}\Int{\d\mathbf{u}}
    e^{-i\q\cdot\mathbf{u}}(f(\mathbf{u}))^{2}
    =\mathcal{O}\left(\frac{1}{d^{4}}\right).
\end{align}

\end{document}